\documentclass[article,12pt]{elsarticle}
\usepackage{amsmath}
\usepackage{float}
\usepackage{comment}
\usepackage{xcolor}

\usepackage{amssymb}

\biboptions{sort&compress}

\usepackage{multirow}
\usepackage{multicol}
\usepackage{graphicx}
\usepackage{verbatim}
\usepackage{comment}
\usepackage{enumerate}
\usepackage{enumitem}
\usepackage{cancel}
\usepackage{xcolor}
\usepackage{bm}

\journal{Applied Mathematics and Computation}

\begin{document}
\begin{frontmatter}


\title{Derivative non-linear Schr\"odinger equation: Singular manifold method and Lie symmetries}

\author{P. Albares$^{a)}$,   P. G. Est\'evez$^{a)}$ and J. D. Lejarreta$^{b)}$}

\address{$^{a)}$ Departamento de F\'{\i}sica Fundamental, Universidad de Salamanca, Salamanca, Spain}

\address{$^{b)}$ Departamento de F\'{\i}sica Aplicada, Universidad de Salamanca, Salamanca, Spain}

\begin{abstract}

We present a generalized study and characterization of the integrability properties of the derivative non-linear Schr\"{o}dinger equation in $1+1$ dimensions. A Lax pair is derived for this equation by means of a Miura transformation and the singular manifold method. This procedure, together with the Darboux transformations, allow us to construct a wide class of rational soliton-like solutions. Lie classical symmetries have also been computed and similarity reductions have been analyzed and discussed. 
\vspace{0.3cm}

\textit{Keywords}: integrability, derivative non-linear Schr\"{o}dinger equation, singular manifold method, Lax pair, Darboux transformations, rational solitons, Lie symmetries, similarity reductions. 
\end{abstract}

\end{frontmatter}

\section{Introduction}

This article is devoted to the study of the derivative non-linear Schr\"{o}dinger (DNLS) equation in $1+1$ dimensions,
\begin{equation}
im_t-m_{xx}-i\left(\left|m\right|^2m\right)_x=0
\label{DNLS0}
\end{equation}
where $m=m(x,t)$ is a complex valued function and the subscripts $x,t$ denote partial derivatives.

DNLS equation has been widely studied in literature in recent years, in terms of integrability characterization, mathematical properties and solutions. This non-linear dispersive equation arises from the field of physics, more specifically in plasma physics and non-linear optics. DNLS equation is found to describe the dynamics of finite-amplitude polarized non-linear Alfv\'en waves propagating parallel to the magnetic field in cold plasma \cite{rogister71,mio1976,mjolhus76} or astrophysical $\beta$-plasma \cite{spangler82,spangler85}. Moreover, DNLS is useful to characterize the behaviour of magnetohydrodynamic (MHD) waves in the Hall-MHD approximation \cite{champeaux99} and also to several amplitude-regimes in different plasma scenarios \cite{kakutani68, ruderman2002a, ruderman2002b,fedun2008}. In the context of optics, DNLS models the propagation of ultra-short pulses in single-mode optical fibers under certain non-linear effects \cite{tzoar81,anderson83,govind01}. 

Besides to its physical relevance and applications, DNLS presents numerous remarkable mathematical properties and analytical solutions of interest. 
Equation \eqref{DNLS0} is an integrable model that may be regarded as a modified version of the famous non-linear  Schr\"{o}\-dinger equation (NLS) \cite{ablow91,drazin89} with a derivative-type non-linearity. 
As NLS equation, DNLS constitutes a non-linear differential equation of reference in the field of mathematical physics and soliton dynamics. Well-posedness of the Cauchy problem for DNLS has been extensively studied in literature and it has been shown that DNLS admits global solutions for a vast range of initial constraints \cite{tsutsumi1980,tsutsumi1981,hayashi1992,takaoka1999,wu2013,wu2015,Pelinovsky2017}. Regarding its integrability properties, in \cite{kaupnew} Kaup and Newell first derived a Lax pair and propose an inverse scattering transform (IST) for DNLS, obtaining the one-soliton solution under the vanishing boundary conditions. Besides, it is also proved in this paper that DNLS equation admits algebraic solitons. Several authors \cite{ichikawa77,wadati78} explored solutions under the non-vanishing boundary conditions and Kawata \textit{et al.} \cite{kawata78,kawata79} investigated the one and two-soliton solution via IST method in this regime. A wide spectrum of mathematical tools, such as the Hirota method \cite{nakamura80,kakei95}, Darboux transformations \cite{imai99, steudel03,xu11}, Hamiltonian formalism and action-angle variables \cite{cai06} or affine Lie groups and symmetry techniques \cite{kakei04,kakei05}, have been applied to analyze DNLS equation. In this context, a plethora of exact soliton-like solutions for DNLS arise: $N$-soliton solution \cite{nakamura80, imai99,steudel03,huang90}, stationary solutions \cite{ichikawa80}, periodic and quasi-periodic solutions \cite{imai99,kamchatnov90,kamchatnov97,kamchatnov00}, breather solutions \cite{ichikawa77,kawata78,xu11,chen04,wang15}, rogue wave solutions \cite{guo13,zhang14,xu19}, etc.

There exit diverse integrable generalizations for this equation, such as multi-component generalizations \cite{morris79,tsuchida99}, extensions to higher dimensions \cite{lou97}, discretized \cite{tsuchida2002,geng2006} or quantized versions of DNLS \cite{kundu1993}. Furthermore, Equation \eqref{DNLS0} is related via gauge transformations \cite{kakei95,wadati83} to several notorious integrable equations, for example the Ablowitz-Kaup-Newell-Segur (AKNS) system \cite{ablowitz74}, or other NLS-like equations with derivative-type non-linearities, as the Chen-Lee-Lie equation \cite{chen79} or the Gerdjikov–Ivanov equation \cite{gerdjikov83}. 
\vspace{0.2cm}

This article is aimed at studying the DNLS equation in $1+1$ dimensions and its associated linear problem. In section \ref{sec:2} the model is presented and its integrability is explored by means of the Painlev\'e analysis \cite{weiss}. Since Painlev\'e test cannot be implemented over DNLS, a previous change of variables and a Miura transformation are required to transform the starting DNLS equation into a suitable system that may have the Painlev\'e property. Besides DNLS, another two PDEs of interest arise naturally through this procedure. In section \ref{sec:3} we shall successfully apply the singular manifold method in order to derive the singular manifold equations and obtain two equivalent Lax pairs for each equation under study. It is worthwhile to remark that this procedure allows us to recover the Lax pair for DNLS proposed by Kaup and Newell \cite{kaupnew}. Binary Darboux transformations and the $\tau$-function formalism is used in section \ref{sec:4} to analyze rational soliton-like solution of DNLS in section \ref{sec:5}. Another different approach for DNLS is conducted in sections \ref{sec:6}-\ref{sec:8}, where Lie classical symmetries, its associated Lie algebra and the similarity reductions are identified and deeply studied. Finally, we close with a section of conclusions. Appendices are introduced at the end of the paper in order to clarify and extend some results concerning the spectral problems of the PDEs involved.

\section{Painlev\'e test  for DNLS: Miura transformations} \label{sec:2}
The derivative non-linear Schr\"odinger equation can be written as the Kaup-Newell system \cite{kaupnew}
\begin{equation}
\begin{aligned}
& im_t-m_{xx}-i\left[\left(m\cdot\overline m\right)m\right]_x=0\\
-& i\overline m_t-\overline m_{xx} + i\left[\left(m\cdot\overline m\right)\overline m\right]_x=0
\end{aligned}
\label{eq0}
\end{equation}
where $\overline m=\overline m(x,t)$ is the complex conjugate of $m=m(x,t)$.

The Painlev\'e test is unable to check the integrability of (\ref{eq0}) because the leading index is not integer. Actually, it is easy to see that the leading index is $-1/2$. This fact allow us to introduce new real fields $\alpha(x,t),\, \beta(x,t)
$ such that
\begin{eqnarray}
&& m=\sqrt{2\alpha_x}\,e^{\frac{i}{2}\beta}\nonumber\\&& \overline m=\sqrt{2\alpha_x}\,e^{-\frac{i}{2}\beta}\label{eq2}\end{eqnarray}

 The introduction of (\ref{eq2}) in the equations (\ref{eq0}) yields
\begin{equation}
\beta=-3\alpha+\int\frac{\alpha_t}{\alpha_x} \,dx\label{eq3}
 \end{equation}
 where $\alpha$ satisfy a differential equation that can be written in conservative form as
 \begin{equation}
 \left[\alpha_x^2-\alpha_t\right]_t=\left[\alpha_{xxx}+\alpha_x^3-\frac{\alpha_t^2+\alpha_{xx}^2}{\alpha_x}\right]_x\label{eq4}
 \end{equation}
 Notice that $\left|m\right|^2=m\cdot \overline{m}=2\alpha_x$ is the density of probability and therefore, $\alpha_x$ is the physically relevant field.
 
 From the point of view of the Painlev\'e property \cite{weiss}, the Painlev\'e test \cite{esther00} can be applied to (\ref{eq4}) due to the fact that the leading terms for $\alpha_x$ and $\alpha_t$ are
 \begin{equation}\alpha_x\sim\pm\,i\,\frac{\phi_x}{\phi}, \quad\quad\alpha_t\sim \pm\,i\,\frac{\phi_t}{\phi} \end{equation}
 where $\phi(x,t)$ is the singular manifold.
 Nevertheless, the existence of two Painlev\'e branches is an inconvenient when the singular manifold method is applied \cite{ecg98}. The restriction to just one  of the two possible signs means that we are loosing a lot of information about equation (\ref{eq4}). According to several previous papers \cite{esther00,estpra04}, the best method to deal with this problem requires the splitting of the field $\alpha$ as
 \begin{equation}\alpha=i(u-\overline u)\label{eq6}\end{equation}
 and according to (\ref{eq4})
  \begin{equation}\alpha_x^2-\alpha_t=u_{xx}+\overline u_{xx}\label{eq7}\end{equation}

 The combination of (\ref{eq6}) and (\ref{eq7}) implies the Miura transformations \cite{ecg98}
\begin{subequations}
\label{eq8}
\begin{equation}
\begin{aligned}
&& u_{xx}=\frac{1}{2}\left(\alpha_x^2-\alpha_t -i\alpha_{xx}\right)\label{eq8a}\end{aligned}\end{equation}\begin{equation}
\begin{aligned}&& \overline u_{xx}=\frac{1}{2}\left(\alpha_x^2-\alpha_t +i\alpha_{xx}\right)\label{eq8b}\end{aligned}\end{equation}\end{subequations} 
as well as a coupling condition between the field $u$ and its complex conjugate $\overline u$, which can be obtained by direct substitution of (\ref{eq6}) in (\ref{eq7}). The result is
\begin{equation}iu_t+u_{xx}-i\overline u_t+\overline u_{xx}+(u_x-\overline u_x)^2=0\label{eq9}\end{equation}
The derivation of equations (\ref{eq8}) with respect to $t$, yields
\begin{subequations}
\label{eq10}
\begin{equation}
\begin{aligned}
&& u_{xt}=\frac{1}{2}\left(\alpha_{xxx}+\alpha_x^3-\frac{\alpha_t^2+\alpha_{xx}^2}{\alpha_x}\right)-\frac{i}{2}\alpha_{xt}\label{eq10a}\end{aligned}\end{equation}\begin{equation}
\begin{aligned}&&\overline u_{xt}= \frac{1}{2}\left(\alpha_{xxx}+\alpha_x^3-\frac{\alpha_t^2+\alpha_{xx}^2}{\alpha_x}\right)+\frac{i}{2}\alpha_{xt}\label{eq10b}\end{aligned}\end{equation}\end{subequations}
where equation (\ref{eq4}) has been used to perform an integration in $x$.

In order to get the equation to be satisfied by $u(x,t)$, we can use (\ref{eq8a}) and (\ref{eq10a}) to obtain
  \begin{eqnarray}\alpha_t&=&\alpha_x^2-i\alpha_{xx}-2u_{xx}\nonumber\\
\alpha_{xxx}&=&-\alpha_x^3+\frac{\alpha_t^2+\alpha_{xx}^2}{\alpha_x}+i\alpha_{xt}+2u_{xt}\label{eq11}\\
\alpha_{xx}&=&
i \left(u_{xx}-\alpha_{x}^2\right) + \frac{\alpha_{x}}{2 u_{xx}}\left(u_{xxx} +iu_{xt}\right)\nonumber\end{eqnarray}

Therefore,  the compatibility condition $ \left(\alpha_t\right)_{xx}=\left(\alpha_{xx}\right)_t$ yields the following non-linear partial differential equation for the field $u(x,t)$ 

\begin{equation}
    \left[u_{tt}+u_{xxxx}+2u_{xx}^2-\frac{u_{xt}^2+u_{xxx}^2}{u_{xx}
}\right]_x=0 \label{eq12}
\end{equation}
Following exactly the same path with equations (\ref{eq8b}) and (\ref{eq10b}), we can easily prove that  $\overline u(x,t)$ should satisfy the same equation (\ref{eq12}). To summarize, $u(x,t)$ and $\overline u(x,t)$ are both solutions of the same PDE (\ref{eq12}) which are also related by the B\"acklund transformation (\ref{eq9}). 

Equation (\ref{eq12}) is known as the non-local Boussinesq equation   \cite{lambert,willox} and its connection to the Kaup system has been extensively studied in \cite{ecg98} from the point of view of the singular manifold method. We are going to use in the next section some of the results of reference \cite{ecg98}.

\section{The Singular Manifold method for DNLS} \label{sec:3}
The advantage of equation (\ref{eq12}) is that besides having the Painlev\'e property, it also has just one Painlev\'e branch \cite{ecg98}. This fact allows us to easily perform the singular manifold method in order to derive many of the properties associated to a non-linear partial differential equation. A list of these properties can be summarized as:
\begin{itemize}

\item The singular manifold equations

\item The Lax pair and its eigenfunctions

\item Darboux transformations of the Lax pair

 \item $\tau$-functions
 
 \item Iterative method for the construction of solutions

\end{itemize}
\subsection{Singular Manifold method}
As it has been proved in \cite{ecg98}, the singular manifold method requires the truncation of the Painlev\'e expansion for $u$ to the constant level (\ref{eq13}), which means that the solutions for  $u$ have to be truncated as 
\begin{equation}u ^{[1]}=\ln(\phi)+u^{[0]}\label{eq13}\end{equation}
This truncation 
acquires the form of an auto-B\"acklund transformation between two solutions $u^{[0]}$ and $u^{[1]}$ of the same equation (\ref{eq12}). Besides that, the manifold $\phi(x,t)$ is not longer an arbitrary function. There are equations (the singular manifold equations) to be satisfied for $\phi$, which can be obtained by direct substitution of (\ref{eq13}) in (\ref{eq12}).  The result read as follows (see \cite{ecg98}):
\begin{itemize}
\item{\bf Expression of the field in terms of the singular manifold}
\begin{subequations}
\label{eq14}
\begin{eqnarray}
&&u^{[0]}_{xx}=-\frac{1}{4}\,\left[v^2+\left(r+2\lambda\right)^2\right]\label{eq14a}\\
&&u^{[0]}_{xt}=\frac{1}{2}\,\left[(r+2\lambda)v_x-vr_x-(r+\lambda)\left(v^2+(r+2\lambda)^2\right)\right]\label{eq14b}
\end{eqnarray}
\end{subequations}
where $\lambda$ is an arbitrary constant and $v$ and $r$ are functions related to the singular manifold through the following definitions
\begin{subequations}
\label{eq15}
\begin{eqnarray}
&& v=\frac{\phi_{xx}}{\phi_x}\label{eq15a}\\
&& r=\frac{\phi_{t}}{\phi_x}\label{eq15b}
\end{eqnarray}
\end{subequations}
\item{\bf Singular Manifold equations}

The equations to be satisfied by the singular manifold could be written as the system
\begin{subequations}
\label{eq16}
\begin{eqnarray}
&&r_t=\left(-v_x+\frac{v^2}{2}+\frac{3\,r^2}{2}+4\lambda r
\right)_x \label{eq16a}\\
&& v_t=\left(r_x+rv\right)_x\label{eq16b}
\end{eqnarray}
\end{subequations}
where equation (\ref{eq16b}) is trivially  obtained from the compatibility condition $(\phi_{xx})_t=(\phi_t)_{xx}$, which arises from the definitions (\ref{eq15a})-(\ref{eq15b}).

It is relevant to remark that  the singular manifold equations are easily related to the Kaup system \cite{kaup}. Actually, we can write system \eqref{eq16} as
\begin{subequations}
\label{eq17}
\begin{eqnarray}
&&\gamma_t=-\eta_{xx}+2\gamma\gamma_x \label{eq17a}\\
&& \eta_t=\gamma_{xx}+2\gamma\eta_x\label{eq17b}
\end{eqnarray}
\end{subequations}
through the following change of variables
\begin{eqnarray}
&&\gamma=r+\lambda\nonumber\\
&&\eta_x=v_x-\frac{v^2}{2}-\frac{(r+2\lambda)^2}{2}\nonumber
\end{eqnarray}
This is an important point. If the singular manifold equations can be considered as the intrinsic canonical form of a PDE, we can conclude that  our original derivative non-linear Schr\"odinger equation (\ref{eq0}) is nothing but a  different form of the Kaup system via Miura transformation. 
\end{itemize}

\subsection{\bf Lax pair}
\begin{itemize}
\item Lax pair for $u(x,t)$

As it was proved in \cite{ecg98}, the singular manifold equations (\ref{eq16}) can be used to to introduce two different  functions $\psi(x,t), \chi(x,t)$ defined as
\begin{subequations}
\label{eq18}
\begin{eqnarray}
&&v=\frac{\psi_x}{\psi}+\frac{\chi_x}{\chi} \label{eq18a}\\
&& r=i\left(\frac{\psi_x}{\psi}-\frac{\chi_x}{\chi}\right)-2\lambda\label{eq18b}
\end{eqnarray}
\end{subequations}
where the term $-2\lambda$ in (\ref{eq18b}) is not essential, but it is useful to simplify the results.

These definitions allow us to linearize  equations (\ref{eq14}). Substitution of (\ref{eq18}) in  the expressions (\ref{eq14}) yields
\begin{subequations}
\begin{eqnarray}\label{eq19}
&& u^{[0]}_{xx}+\frac{\psi_x}{\psi}\frac{\chi_x}{\chi} =0\label{eq19a}\\
&& u^{[0]}_{xt}+\frac{\psi_x}{\psi}\frac{\chi_x}{\chi} \left(i\frac{\psi_{xx}}{\psi_x}-i\frac{\chi_{xx}}{\chi_x}+i\frac{\psi_{x}}{\psi}-i\frac{\chi_{x}}{\chi}-2\lambda\right)=0\label{eq19b}
\end{eqnarray}
\end{subequations}
These equations can easily combined in order to obtain
\begin{subequations}
\label{eq20}
\begin{eqnarray}
&& \psi_{xx}=\left(\frac{u^{[0]}_{xxx}-iu^{[0]}_{xt}}{2u^{[0]}_{xx}}-i\lambda\right)\,\psi_x-u^{[0]}_{xx}\psi\label{eq20a}\\
&& \chi_{xx}=\left(\frac{u^{[0]}_{xxx}+iu^{[0]}_{xt}}{2u^{[0]}_{xx}}+i\lambda\right)\,\chi_x-u^{[0]}_{xx}\chi\label{eq20b}
\end{eqnarray}
\end{subequations}
Besides that, the singular manifold equations (\ref{eq16}) provide

\begin{eqnarray}
&& \frac{\psi_t}{\psi}+\frac{\chi_t}{\chi}-i\left(\frac{\psi_{xx}}{\psi}-\frac{\chi_{xx}}{\chi}\right)+2\lambda\left(\frac{\psi_x}{\psi}+\frac{\chi_x}{\chi}\right)=0\nonumber\\
&&i \left(\frac{\psi_t}{\psi}-\frac{\chi_t}{\chi}\right)
+\left(\frac{\psi_{xx}}{\psi}+\frac{\chi_{xx}}{\chi}\right)+2i\lambda\left(\frac{\psi_x}{\psi}-\frac{\chi_x}{\chi}\right)+4u^{[0]}_{xx}+2\lambda^2=0\nonumber
\end{eqnarray}

The combination of these two last equations allows us to write
\begin{subequations}
\label{eq21}
\begin{eqnarray}
&& \psi_t=i\psi_{xx}-2\lambda\psi_x+i\left(2u^{[0]}_{xx}+\lambda^2\right)\psi\label{eq21a}\\
&&\chi_t=-i\chi_{xx}-2\lambda\chi_x-i\left(2u^{[0]}_{xx}+\lambda^2\right)\chi\label{eq21b}
\end{eqnarray}
\end{subequations}
Equations (\ref{eq20}) and (\ref{eq21}) are a Lax pair for equation (\ref{eq12}). Notice that the two eigenfuntions are also related by (\ref{eq19a}). Definitions (\ref{eq18}) can be easily combined with (\ref{eq15}) to provide the singular manifold through the exact derivative
\begin{equation}d\phi=\psi\,\chi dx+\left[-2\lambda \psi\,\chi+i\left(\chi\,\psi_x-\psi\,\chi_x\right)\right]dt\label{eq22}\end{equation}
whose integration allows us to obtain the iterated solution (\ref{eq13}).
Obviously, the complex conjugate form of (\ref{eq20}) and (\ref{eq21}) is the Lax pair for $\overline u^{[0]}(x,t)$.

\item{Lax pair for $\alpha(x,t)$}

We can now use our previous results to derive a Lax pair for the field $\alpha(x,t)$. Combination of (\ref{eq6}) and the Painlev\'e expansion (\ref {eq13}) gives a Painlev\'e expansion for $\alpha $ of the form
\begin{equation}\alpha^{[1]}=i\left(u^{[1]}-\overline u^{[1]}\right)=\alpha^{[0]}+i\ln\left(\frac{\phi}{\overline \phi}\right)\label{eq23}\end{equation}
where $\overline \psi$ and $\overline \chi $ should be introduced as the complex conjugates of $\psi$ and $\chi$ in order to have the complex conjugate of (\ref{eq22}). Besides that, the coupling condition (\ref{eq9}) should be fulfilled for $u^{[0]}$ and $\overline u^{[0]}$. It imposes an additional condition for the singular manifold $\phi(x,t)$ and its complex conjugate $\overline \phi(x,t)$ that can be written as
\begin{equation}\frac{\phi_x}{\phi}\left(\frac{\overline \chi_x}{\overline \chi}+i\alpha^{[0]}_x+i\lambda\right)+\frac{\overline \phi_x}{\overline\phi}\left(\frac{ \chi_x}{\chi}-i\alpha^{[0]}_x-i\lambda\right)=1\end{equation}
A Lax pair for $\alpha^{[0]} $ can be easily obtained from  (\ref{eq20b}) and (\ref{eq21b}) with the aid of (\ref{eq8}) and (\ref{eq10}). The result is
\begin{eqnarray}
&& \chi_{xx}=i\left[\frac{\alpha^{[0]}_t+\left(\alpha^{[0]}_x\right)^2-i\alpha^{[0]}_{xx}}{2\alpha^{[0]}_x}+\lambda\right]\chi_x+\left[\frac{\alpha^{[0]}_t-\left(\alpha^{[0]}_x\right)^2+i\alpha^{[0]}_{xx}}{2}\right]\chi\nonumber\\
&&\chi_t=-i\chi_{xx}-2\lambda\chi_x+i\left[\alpha^{[0]}_t-\left(\alpha^{[0]}_x\right)^2+i\alpha^{[0]}_{xx}-\lambda^2\right]\chi\label{eq25}
\end{eqnarray}
An alternative Lax pair can be obtained by substitution of (\ref{eq8}) and (\ref{eq10}) in (\ref{eq20a}) and (\ref{eq21a}). We omit here its explicit expression, because the Lax pair in terms of $\psi$ can be easily obtained by using  (\ref{eq19a}) in (\ref{eq25}). We may refer the reader to the Appendix \ref{alphaLP} for the complete expression this alternative Lax pair.
\item{Lax pair for $m(x,t)$}

Finally, from equations (\ref{eq2}) and (\ref{eq3}), we can obtain the derivatives of $\alpha $ as
\begin{eqnarray}
&&\alpha^{[0]}_x=\frac{m^{[0]}\overline m^{[0]}}{2}\nonumber\\&&\alpha^{[0]}_t= \frac{3}{4}\left(m^{[0]}\right)^2\left(\overline m^{[0]}\right)^2+\frac{i}{2}\left(m^{[0]}\overline m^{[0]}_x-\overline m^{[0]} m^{[0]}_x\right)\label{eq26}
\end{eqnarray}
which allows us to write the Lax pair (\ref{eq25}) for the initial system (\ref{eq0}) as
\begin{eqnarray}
&& \chi_{xx}=i\left[m^{[0]}\overline m^{[0]}-i\frac{m^{[0]}_x}{m^{[0]}}+\lambda\right]\chi_x+\frac{m^{[0]}\overline m^{[0]}}{2}\left[\frac{m^{[0]}\overline m^{[0]}}{2}+i\frac{\overline m^{[0]}_x}{\overline m^{[0]}}\right]\chi\nonumber\\
&&\chi_t=-i\chi_{xx}-2\lambda\chi_x+i\left[m^{[0]}\overline m^{[0]}\left(\frac{m^{[0]}\overline m^{[0]}}{2}+i\frac{\overline m^{[0]}_x}{\overline m^{[0]}}\right)-\lambda^2\right]\chi\label{eq27}
\end{eqnarray}

Analogously, an alternative Lax pair for $m(x,t)$ may be found in Appendix \ref{LPDNLS1}. 
\end{itemize}

\section{Darboux Transformations}\label{sec:4}
As it has been shown in many previous articles \cite{esther00,estpra04,villarroel1,estevez2,albares,diaz,ecg98}, once the Lax pair have been obtained for a given PDE by means of the singular manifold method, a binary Darboux transformation can be constructed. The method can be described as follows:
\subsection{Seed eigenfuctions}

Let be $\chi_i,\, i=1,2$ two different eigenfunctions of the Lax pairs (\ref{eq20a})-(\ref{eq21a}) for the seed field $u^{[0]}$, corresponding to two different eigenvalues $\lambda_i,\, i=1,2$. Therefore, we have
\begin{subequations}
\label{eq28}
\begin{eqnarray}
&& (\chi_i)_{xx}=\left(\frac{u^{[0]}_{xxx}+iu^{[0]}_{xt}}{2u^{[0]}_{xx}}+i\lambda_i\right)\,(\chi_i)_x-u^{[0]}_{xx}\,\chi_i\\
&& (\chi_i)_t=-i (\chi_i)_{xx}-2\lambda_i(\chi_i)_x-i\left(2u^{[0]}_{xx}+\lambda_i^2\right)\,\chi_i
\end{eqnarray}
\end{subequations}
Furthermore, two different eigenfunctions $\psi_i,\, i=1,2$ can be introduced by means of (\ref{eq19a}) in the form 
h
\begin{equation}u^{[0]}_{xx}+\frac{(\psi_i)_x}{\psi_i}\frac{(\chi_i)_x}{\chi_i} =0\end{equation}
which means that we can introduce two different singular manifolds $\phi_i,\, i=1,2$ defined in (\ref{eq22}) through the expression

\begin{equation}d\phi_i=\psi_i\,\chi_i\, dx+\left\{-2\lambda_i \psi_i\,\chi_i+i\left[\chi_i\,(\psi_i)_x-\psi_i\,(\chi_i)_x\right]\right\}dt\label{eq30}\end{equation}
\subsection{Iterated eigenfunctions}
As we have seen in the previous section, the truncated Painlev\'e expansion (\ref{eq13}) can be considered as an auto-B\"acklund transformation
\begin{equation}u^{[1]}=\ln(\phi_1)+u^{[0]}\label{eq31}\end{equation}
which allows us to obtain an iterated field $u^{[1]}$. 
Obviously an iterated Lax pair can be defined for this iterated field in the form
\begin{subequations}
\label{eq32}
\begin{eqnarray}
&& (\chi_{1,2})_{xx}=\left(\frac{u^{[1]}_{xxx}+iu^{[1]}_{xt}}{2u^{[1]}_{xx}}+i\lambda_2\right)\,(\chi_{1,2})_x-u^{[1]}_{xx}\,(\chi_{1,2})\\
&& (\chi_{1,2})_t=-i (\chi_{1,2})_{xx}-2\lambda_2(\chi_{1,2})_x-i\left(2u^{[1]}_{xx}+\lambda_2^2\right)\,(\chi_{1,2})\\
&&u^{[1]}_{xx}+\frac{(\psi_{1,2})_x}{\psi_{1,2}}\frac{(\chi_{1,2})_x}{\chi_{1,2}} =0
\end{eqnarray}
\end{subequations}
and consequently, a  singular manifold $\phi_{1,2}$ can be defined for the iterated field though the expression
\begin{equation}d\phi_{1,2}=\psi_{1,2}\,\chi_{1,2}\, dx+\left\{-2\lambda_2 \psi_{1,2}\,\chi_{1,2}+i\left[\chi_{1,2}\,(\psi_{1,2})_x-\psi_{1,2}\,(\chi_{1,2})_x\right]\right\}dt\label{eq33}\end{equation}

\subsection{Second Iteration}
The singular manifold $\phi_{1,2}$, as defined in (\ref{eq33}), can be used to perform a second iteration such that a new field $u^{[2]}$ can be constructed as 
\begin{equation}u^{[2]}=\ln(\phi_{1,2})+u^{[1]}\label{eq34}\end{equation}
A Lax pair, is usually considered system of equations which is linear in the eigenfuctions. A different point of view \cite{esther00,ecg98}, is the consideration of the equations (\ref{eq32}) as non-linear relations between the field $u^{[1]}$ and the eigenfunction $\chi_{1,2}$.  According to this new consideration, the Painlev\'e expansion (\ref{eq31}) for the field, should be accompanied by a similar expansion for the eigenfunctions. Let be
\begin{subequations} 
\label{eq35}
\begin{eqnarray} &&\chi_{1,2}=\chi_2-\chi_1\,\frac{\Delta_{1,2}}{\phi_1}\\
&&
\psi_{1,2}=\psi_2-\psi_1\,\frac{\Sigma_{1,2}}{\phi_1}
\end{eqnarray}\end{subequations}
such  expansion. $\Delta_{1,2}$ and $\Sigma_{1,2}$ are functions to be determined later.
Besides that, (\ref{eq33}) implies that we can also provide a Painlev\'e expansion for the singular manifold in the form

\begin{equation} \phi_{1,2}=\phi_2+\frac{\Omega_{1,2}}{\phi_1}\label{eq36}\end{equation}
Substitution of (\ref{eq31}),  (\ref{eq35})  and (\ref{eq36})  in (\ref{eq32})  and (\ref{eq33}) requires a lot of calculation, which is not difficult to handle with the symbolic package Maple. The final results read as
\begin{subequations}
\label{eq37}
\begin{eqnarray}
&& \Sigma_{1,2}=\Delta_{2,1}\label{eq37a}\\
&&\Omega_{1,2}=-\Delta_{1,2}\Delta_{2,1}\label{eq37b}\\
&&\Delta_{i,j}=i\psi_i\,\frac{\chi_i(\chi_j)_x-\chi_j(\chi_i)_x}{\left(\lambda_i-\lambda_j\right)(\chi_i)_x}\label{eq37c}
\end{eqnarray}
\end{subequations}
Therefore, we can conclude that  (\ref{eq31}) and  (\ref{eq35})  are binary Darboux transformations for the Lax pair  (\ref{eq32}). The  eigenfunctions $\chi_i$ and $\psi_i$ for the seminal solution $u^{[0]}$ are the only tools which we need in order to construct the iterated solution (\ref{eq31}).
\subsection{$\tau$-function}
By combining (\ref{eq31}) and (\ref{eq34}), we can get the second iterated solution as 
\begin{equation}u^{[2]}=\ln(\phi_1\phi_{1,2})+u^{[0]}\label{eq38}\end{equation}
By using (\ref{eq36}) and (\ref{eq37b}), (\ref{eq38}) can be written as
\begin{eqnarray}&&u^{[2]}=\ln(\tau_{1,2})+u^{[0]}\nonumber\\
&&\tau_{1,2}=\phi_1\phi_2-\Delta_{1,2}\Delta_{2,1}\label{eq39}\end{eqnarray}

This procedure may be implemented repeatedly and generalized up to the $n$th-iteration, which read as
\begin{equation}
    u^{[n]}=\ln(\phi_1\phi_{1,2}\cdots\phi_{1,2,\ldots,n})+u^{[0]}=\ln(\tau_{1,2,\ldots,n})+u^{[0]}
\end{equation}

The $\tau$-function for the $n$th-iteration can be computed as
\begin{equation}
    \tau_{1,2,\ldots,n}=\operatorname{det}\left(\Delta_{i,j}\right)_n,\qquad i,j=1,\ldots, n
\end{equation}

where $\left(\Delta_{i,j}\right)_n$ denotes the $n\times n$ matrix of entries
\begin{equation}
    \left\{\begin{array}{lll}
       \Delta_{i,i}=\phi_i  & \text{for} & i=j\\
       \Delta_{i,j}=i\psi_i\,\frac{\chi_i(\chi_j)_x-\chi_j(\chi_i)_x}{\left(\lambda_i-\lambda_j\right)(\chi_i)_x}  & \text{for} & i\neq j
    \end{array}\right.
\end{equation}

that may be exclusively expressed in terms of $n$ different couples of eigenfunctions $\{\chi_k,\psi_k\}$ of eigenvalues $\lambda_k$, for the seminal Lax pairs \eqref{eq20}-\eqref{eq21} and $n$ singular manifolds $\phi_k$ given by \eqref{eq30}, $k=1,\ldots,n$.

\section{Rational solitons}\label{sec:5}
In this section, rational soliton-like solutions for the DNLS equation are obtained by applying the procedure described above. The density of probability for the DNLS equations, the relevant physical field associated to the formation of solitons, may be expressed as
\begin{equation}
    \left|m\right|^2=m\cdot \overline{m}=2i\left(u_x-\overline{u}_x\right)
    \label{eq43}
\end{equation}
We will start with a seed solution $u^{[0]}$ and a couple of eigenfunctions for its Lax pair $\{\chi_j,\psi_j\},\,j=1,2$ that allow us to construct the first and the second iteration, $u^{[1]}$ and $u^{[2]}$, the $\Delta$ matrix and $\tau$-function, which will lead straightforward to the soliton solution profile.

Let us consider the following seed solution for \eqref{eq0},
\begin{equation}
    m^{[0]}=j_0\,e^{\frac{i}{2}j_0^2(z_0^2-1)\left[x+\frac{j_0^2}{2}\left(z_0^2+1\right)t\right]}
    \label{seedm}
\end{equation}
where $j_0,\,z_0$ are arbitrary constants. This seed solution leads to a polynomial solution in $u$ and $\overline{u}$ for \eqref{eq12} as
\begin{equation}
\begin{aligned}
u^{[0]}&=-\frac{j_0^2}{4}\left[j_0^2z_0^2x\left(\frac{x}{2}+j_0^2(z_0^2+1)\,t\right)+i\left(x+j_0^2\left(z_0^2+\frac{1}{2}\right)t\right)\right]\\
\overline{u}^{[0]}&=-\frac{j_0^2}{4}\left[j_0^2z_0^2x\left(\frac{x}{2}+j_0^2(z_0^2+1)\,t\right)-i\left(x+j_0^2\left(z_0^2+\frac{1}{2}\right)t\right)\right]
\end{aligned}
\end{equation}
where condition \eqref{eq9} is identically satisfied.

Solutions for the Lax pair \eqref{eq20}-\eqref{eq21} may be constructed as
\begin{equation}
    \begin{aligned}
    \chi_{\sigma}&=\,e^{\,\,\frac{i}{2}j_0^2z_0\sigma\,\left[x+j_0^2\left(-\frac{\sigma}{2z_0}(z_0^4+7z_0^2+1)+3(z_0^2+1)\right)t\right]}\\
    \psi_{\sigma}&=e^{-\frac{i}{2}j_0^2z_0\sigma\left[x+j_0^2\left(-\frac{\sigma}{2z_0}(z_0^4+7z_0^2+1)+3(z_0^2+1)\right)t\right]}
    \label{eq46}
    \end{aligned}
\end{equation}
where these eigenfunctions depend on an additional binary real parameter $\sigma$, such that $\sigma^2=1$. 
The spectral parameter associated to these eigenfunctions is written as
\begin{equation}
    \lambda_{\sigma}=\frac{j_0^2}{2}\left(2\sigma z_0-(z_0^2+1)\right)
    \label{eq47}
\end{equation}

By means of equation \eqref{eq30}, we get the singular manifold, which also depends on $\sigma$,
\begin{equation}
    \phi_{\sigma}=x-j_0^2\left(\sigma z_0-(z_0^2+1)\right)t-\frac{i}{j_0^2z_0(\sigma-z_0)}
    \label{eq48}
\end{equation}
\subsection{First iteration and one-soliton solution}
Then, it is possible to compute the first iteration through \eqref{eq31}, 
\begin{equation}
\begin{aligned}
    u^{[1]}_{\sigma}=&-\frac{j_0^2}{4}\left[j_0^2z_0^2x\left(\frac{x}{2}+j_0^2(z_0^2+1)\,t\right)+i\left(x+j_0^2\left(z_0^2+\frac{1}{2}\right)t\right)\right]\\
    &+\ln{\left(x-j_0^2\left(\sigma z_0-(z_0^2+1)\right)t-\frac{i}{j_0^2z_0(\sigma-z_0)}\right)}\\
    \overline{u}^{[1]}_{\sigma}=&-\frac{j_0^2}{4}\left[j_0^2z_0^2x\left(\frac{x}{2}+j_0^2(z_0^2+1)\,t\right)-i\left(x+j_0^2\left(z_0^2+\frac{1}{2}\right)t\right)\right]\\
    &+\ln{\left(x-j_0^2\left(\sigma z_0-(z_0^2+1)\right)t+\frac{i}{j_0^2z_0(\sigma-z_0)}\right)}
    \end{aligned}
\end{equation}
where we can check that $u_{\sigma}^{[1]}$ and $\overline{u}_{\sigma}^{[1]}$ are complex conjugates. 

Hence, the density of probability for the first iteration is deduced from \eqref{eq43} as
\begin{equation}
    \left|m^{[1]}_{\sigma}\right|^2=\left|m^{[0]}_{\sigma}\right|^2-4\operatorname{Im}\left(\frac{(\phi_{\sigma})_x}{\phi_{\sigma}}\right)=j_0^2-\frac{4}{j_0^2z_0(\sigma-z_0)\left[\left(x-v_{\sigma}t\right)^2+\frac{1}{j_0^4z_0^2(\sigma-z_0)^2}\right]}
    \label{eq50}
\end{equation}
which corresponds to a travelling rational soliton-like wave along the $x-v_{\sigma}t$ direction, of speed
\begin{equation}
    v_{\sigma}=j_0^2\left(\sigma z_0-(z_0^2+1)\right)
    \label{eq51}
\end{equation}
and constant amplitude
\begin{equation}
    a_{\sigma}=-j_0^2\left(4z_0(\sigma-z_0)-1\right)
\end{equation}
One may observe that depending on the values of the parameters $\sigma=\pm 1$ and $z_0$ is possible to obtain either bright or dark rational solitons.

\begin{figure}[H]
    \centering
    \includegraphics[width=0.325\textwidth]{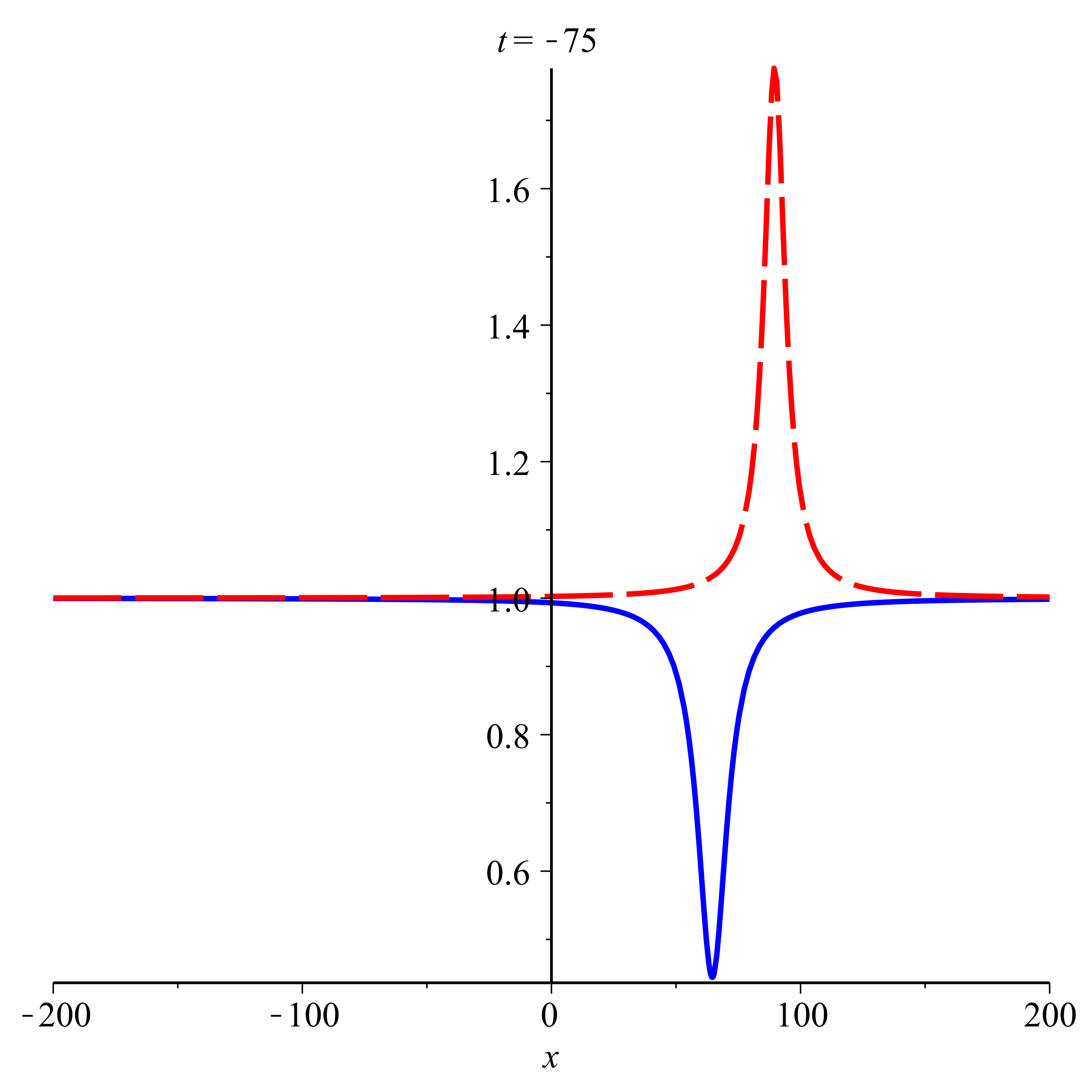}
    \includegraphics[width=0.325\textwidth]{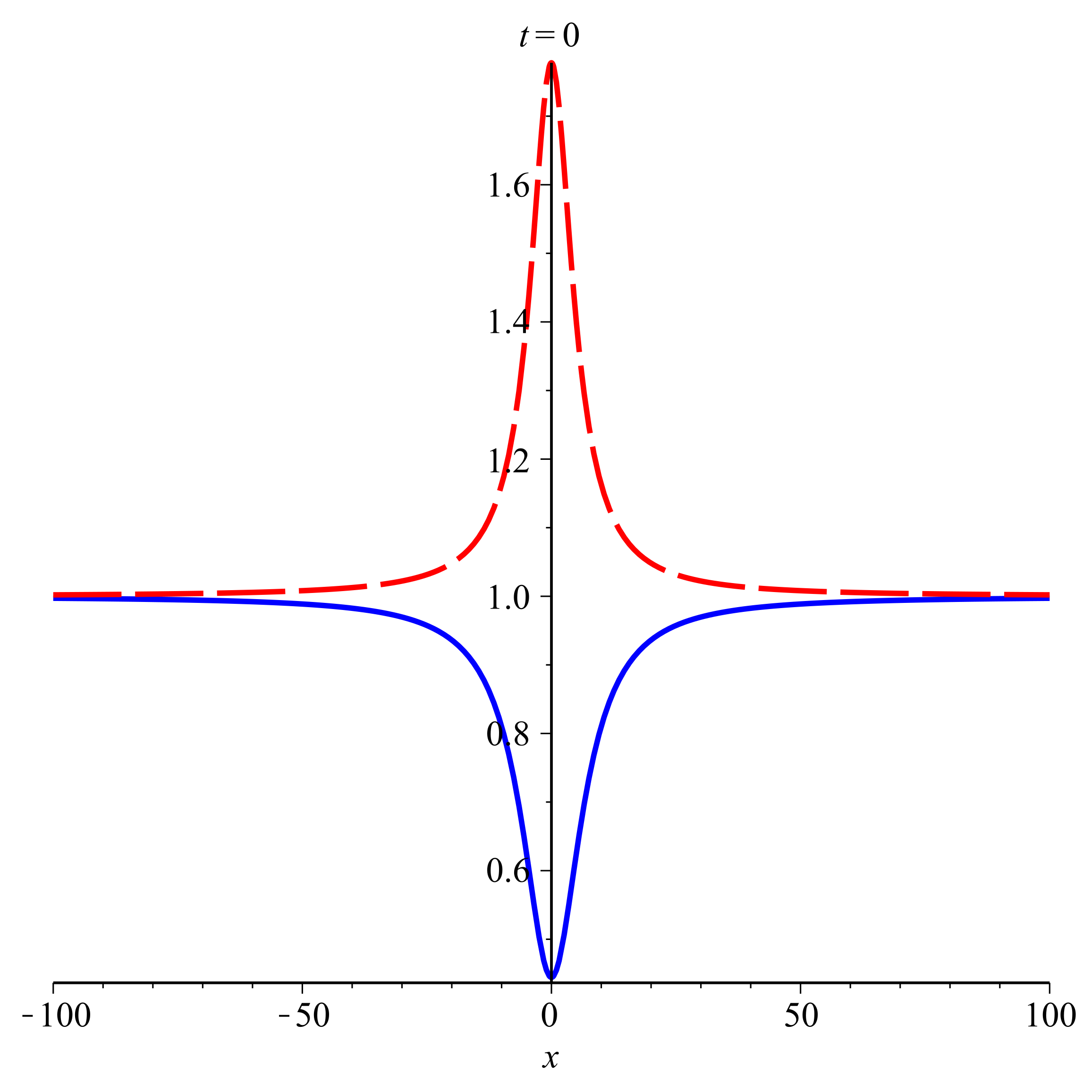}
    \includegraphics[width=0.325\textwidth]{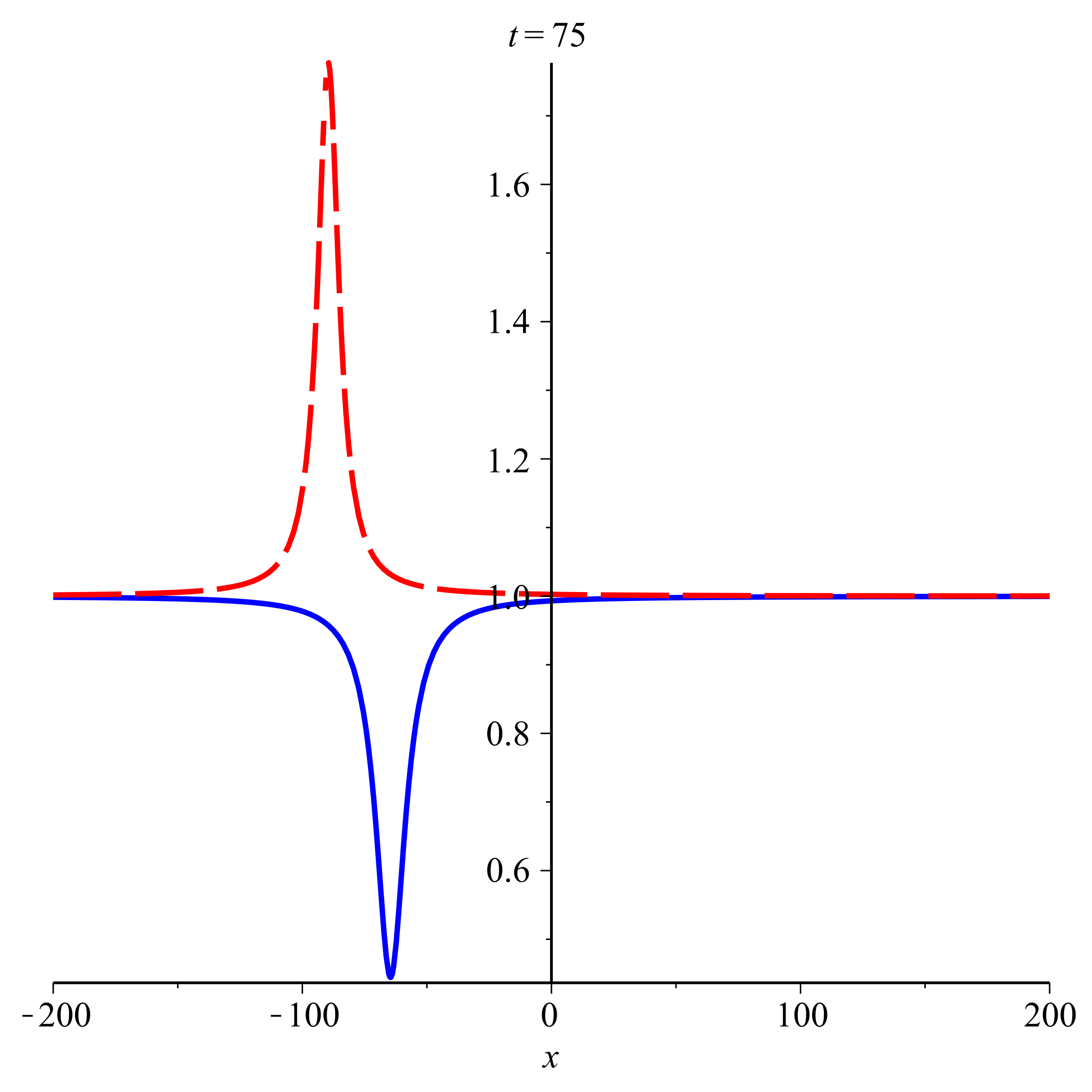}
    \caption{One rational soliton solutions $\left|m^{[1]}_{\sigma}\right|^2$ at times $t=-75,\,0,\,75$. The solid blue line represents a dark soliton for $\sigma=1,\,j_0=1,\,z_0=\frac{1}{6}$, and the dashed red line displays a bright soliton for $\sigma=-1,\,j_0=1,\,z_0=\frac{1}{6}$.}
    \label{fig1}
\end{figure}

These one soliton solutions $\left|m^{[1]}_{\sigma}\right|^2$ are displayed in Figure \ref{fig1} at different times, where a bright rational soliton is obtained for $\sigma=-1$ and a dark one for $\sigma=1$.
\subsection{Second iteration and two-soliton solution}
In order to perform the second iteration and obtain the two-soliton solution, we need to compute two different sets of eigenfunctions $\{\chi_j,\psi_j\}$ with eigenvalues $\lambda_j$ that lead to the singular manifolds $\phi_j$, with $j=1,2$. Thus, in the following, we will identify the first set of functions $\{\chi_1,\,\psi_1,\,\lambda_1,\,\phi_1\}$ as those with $\sigma=1$ in equations \eqref{eq46}-\eqref{eq48}; and the second set, $\{\chi_2,\,\psi_2,\,\lambda_2,\,\phi_2\}$, with $\sigma=-1$, respectively.

After this identification, it is immediate to obtain the $\Delta$ matrix and the $\tau$-function through equations \eqref{eq37c} and \eqref{eq39}, of final expressions
\begin{equation}
    \Delta_{1,2}=-\frac{i}{j_0^2z_0}\,e^{-ij_0^2z_0\,\left[x+3j_0^2\left(z_0^2+1\right)\,t\,\right]},\qquad \Delta_{2,1}=\,\frac{i}{j_0^2z_0}\,e^{ij_0^2z_0\,\left[x+3j_0^2\left(z_0^2+1\right)\,t\,\right]}
\end{equation}
\begin{equation}
    \tau_{1,2}=\left(x+j_0^2(z_0^2+1)\,t\right)^2-j_0^4z_0^2t^2+\frac{2ij_0^2\left(x+j_0^2(z_0^2+2)\,t\right)-1}{j_0^4(z_0^2-1)}
\end{equation}
Therefore, the second iteration for $u$ and $\overline u$ is given by \eqref{eq39}
\begin{equation}
\begin{aligned}
    u^{[2]}=&-\frac{j_0^2}{4}\left[j_0^2z_0^2x\left(\frac{x}{2}+j_0^2(z_0^2+1)\,t\right)+i\left(x+j_0^2\left(z_0^2+\frac{1}{2}\right)t\right)\right]\\[.2cm]
    &+\ln{\left(\left(x+j_0^2(z_0^2+1)\,t\right)^2-j_0^4z_0^2t^2+\frac{2ij_0^2\left(x+j_0^2(z_0^2+2)\,t\right)-1}{j_0^4(z_0^2-1)}\right)}\\
    \overline{u}^{[2]}=&-\frac{j_0^2}{4}\left[j_0^2z_0^2x\left(\frac{x}{2}+j_0^2(z_0^2+1)\,t\right)-i\left(x+j_0^2\left(z_0^2+\frac{1}{2}\right)t\right)\right]\\[.2cm]
    &+\ln{\left(\left(x+j_0^2(z_0^2+1)\,t\right)^2-j_0^4z_0^2t^2-\frac{2ij_0^2\left(x+j_0^2(z_0^2+2)\,t\right)+1}{j_0^4(z_0^2-1)}\right)}
\end{aligned}    
\end{equation}
and $ \left|m^{[2]}\right|^2$ acquires the final form
\small
\begin{equation}
\begin{aligned}
    \left|m^{[2]}\right|^2&=\left|m^{[0]}\right|^2-4\operatorname{Im}\left(\frac{(\tau_{1,2})_x}{\tau_{1,2}}\right)=j_0^2\\
    &+\frac{8\left[\left(x+j_0^2(z_0^2+2)\,t\right)^2+j_0^4(z_0^2-1)\,t^2+\frac{1}{j_0^4(z_0^2-1)}\right]}{j_0^2(z_0^2-1)\left[\left(\left(x+j_0^2(z_0^2+1)\,t\right)^2-j_0^4z_0^2t^2-\frac{1}{j_0^4(z_0^2-1)}\right)^2+\frac{4\left(x+j_0^2(z_0^2+2)\,t\right)^2}{j_0^4(z_0^2-1)^2}\right]}
\end{aligned}
\end{equation}
\normalsize
\subsection{Asymptotic behaviour}
This solution asymptotically yields two rational solitons moving along the lines $x-v_{\sigma}t$ of the form \eqref{eq50}, with speed \eqref{eq51} for $\sigma=\pm 1$, respectively. In order to enlighten this point, an asymptotic behaviour for each rational soliton may be performed. Let us consider the following transformation
\begin{equation}
    X_1=x-v_1t,\qquad v_1=-j_0^2\left(z_0^2-z_0+1\right)
\end{equation}
that allow us to write the limit of $ \left|m^{[2]}\right|^2$ at $t\to\pm \infty$ as the static rational soliton
\begin{equation}
    \left|m^{[2]}\right|^2\sim j_0^2+\frac{4}{j_0^2z_0(z_0-1)\left[X_1^2+\frac{1}{j_0^4z_0^2(z_0-1)^2}\right]}
\end{equation}
which correspond to the first iteration solution \eqref{eq50} for $\sigma=1$. 

A complete analogous analysis can be consider for the second soliton, by means of the transformation
\begin{equation}
\begin{aligned}
    X_2&=x-v_2t,\qquad v_2=-j_0^2\left(z_0^2+z_0+1\right)\\
    \left|m^{[2]}\right|^2&\sim j_0^2+\frac{4}{j_0^2z_0(z_0+1)\left[X_2^2+\frac{1}{j_0^4z_0^2(z_0+1)^2}\right]}
    \end{aligned}
\end{equation}
that leads to a similar profile for \eqref{eq50} with $\sigma=-1$.

\begin{figure}[H]
    \centering
    \includegraphics[width=0.325\textwidth]{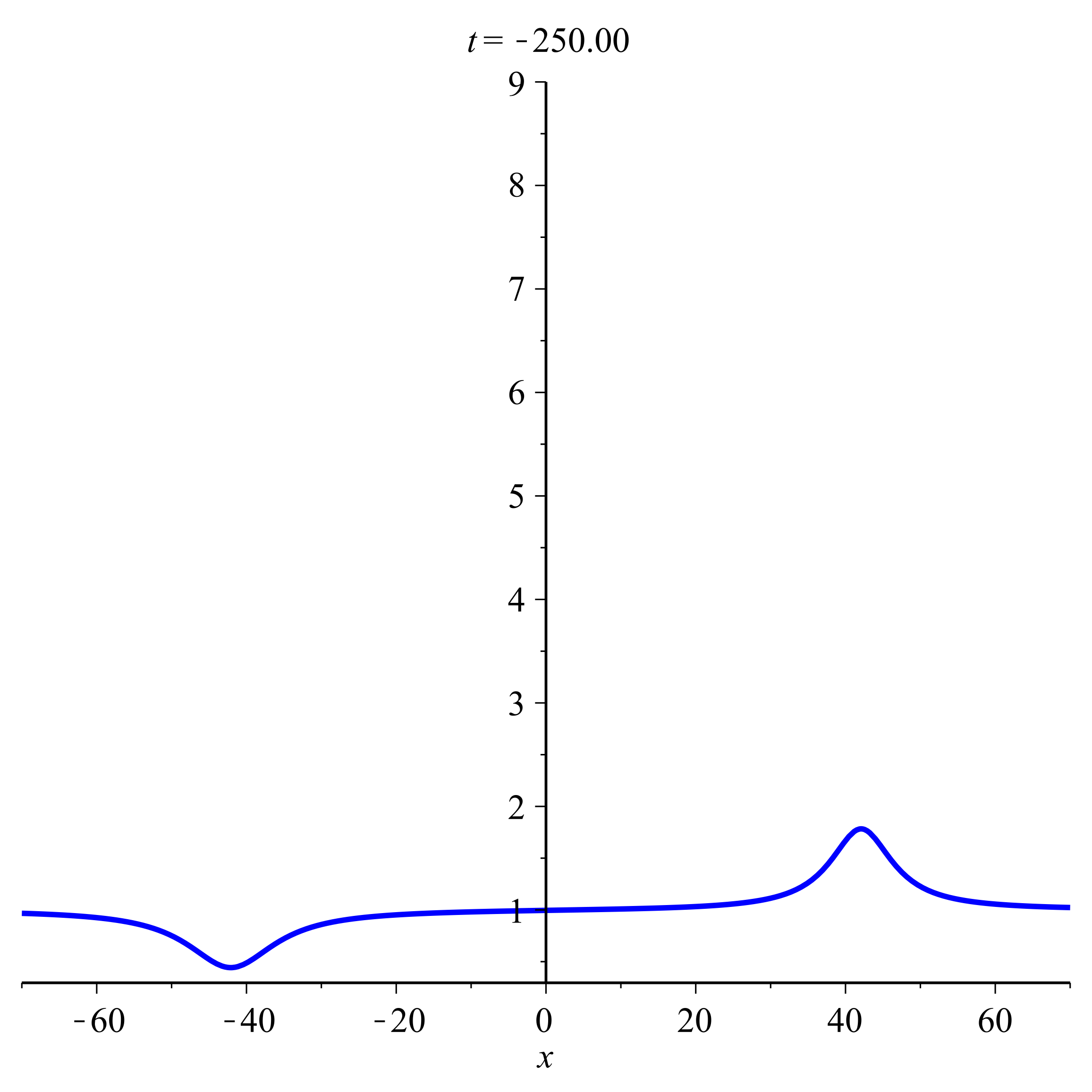}
    \includegraphics[width=0.325\textwidth]{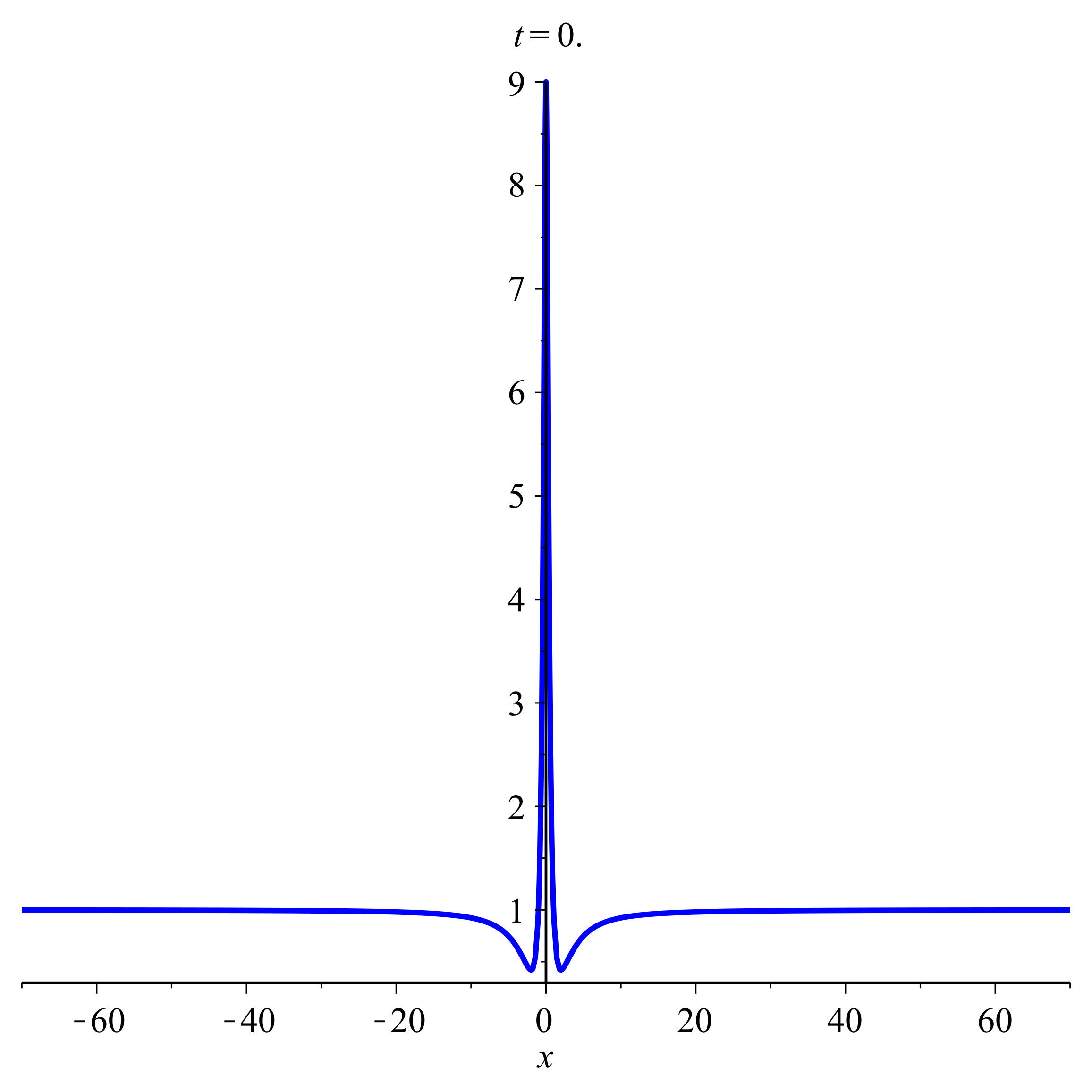}
    \includegraphics[width=0.325\textwidth]{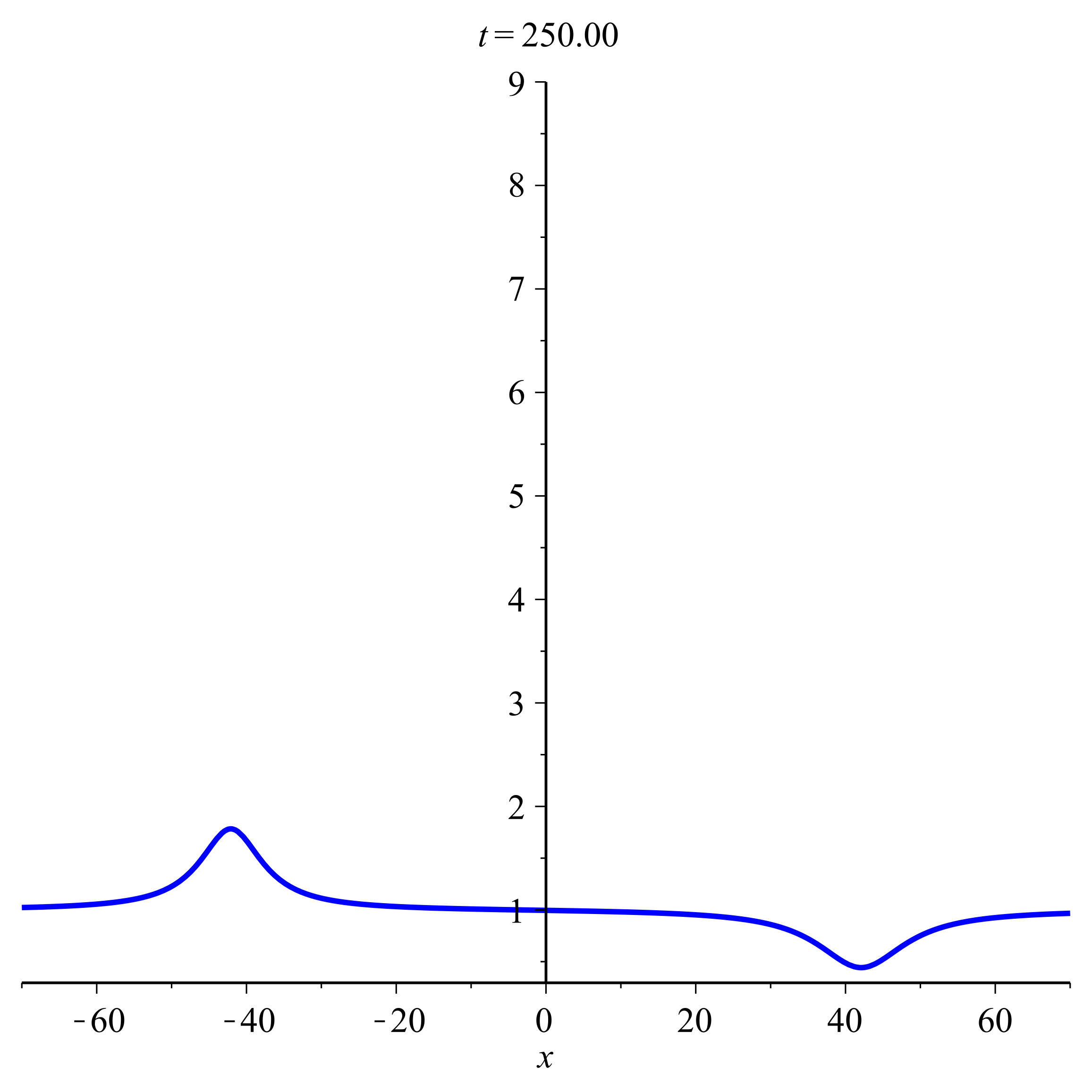}
    \caption{Two rational soliton solution $\left|m^{[2]}\right|^2$ at times $t=-250,\,0,\,250$, for $j_0=1,\,z_0=\frac{1}{6}$.}
    \label{fig2}
\end{figure}

\begin{figure}[H]
    \centering
    \includegraphics[width=0.6\textwidth]{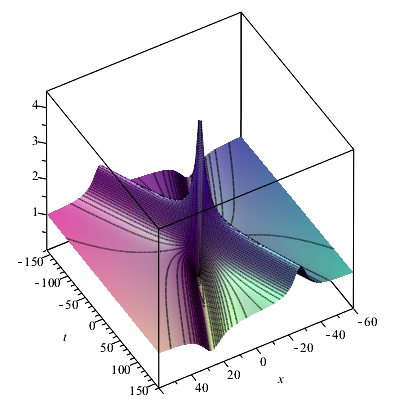}
    \caption{Spatio-temporal plot of the two rational soliton solution $\left|m^{[2]}\right|^2$ for parameters $j_0=1,\,z_0=\frac{1}{6}$.}
    \label{fig3}
\end{figure}

Figure \ref{fig2} displays the two-soliton solution $\left|m^{[2]}\right|^2$ at different times. In Figure \ref{fig3}, a 3D spatio-temporal plot of the two-soliton solution is also presented. Both figures \ref{fig2} and \ref{fig3} have been plotted in the center-of-mass reference frame of the two colliding rational solitons, which may be achieved after the galilean transformation $x=X_{\text{CM}}+\frac{1}{2}\left(v_1+v_2\right)t$. In this system of reference, the two rational solitons move with equal and opposite velocities $c=\frac{1}{2}\left(v_1-v_2\right)$ along the lines $X_{\text{CM}}-\sigma ct$.
In these graphics the scattering between the bright and the dark rational solitons is explicitly appreciated. It is worthwhile to emphasize the unperturbed asymptotic behaviour of both solitons.

\section{Classical Lie symmetries}\label{sec:6}

This section is devoted to the Lie symmetry analysis. Instead of the Lie symmetries for the equation, we are dealing with the corresponding ones for all Lax pairs given above, Lax pairs \eqref{eq20}-\eqref{eq21} for $u$, Lax pair \eqref{eq25} for $\alpha$ and Lax pair \eqref{eq27} for $m$, together with the correspondent alternative Lax pairs for $\psi$, written in \ref{app:LP}, respectively. Theoretical background about the classical Lie method to compute Lie symmetries may be found in the textbooks ~\cite{BK,Lie,Olver,Steph}.

In order to compute correctly the Lie symmetries, it is necessary to consider both $\chi$-Lax pairs and $\psi$-Lax pair, together with the $\{\chi,\psi\}$-coupling condition \eqref{eq19a}, and their complex conjugates. With the purpose of illustrating a general framework of Lie's method, valid for each case under consideration, let us define a generic field $\Omega=\Omega(x,t)$ that may act as $\{m,\alpha,u\}$ and its complex conjugate $\overline{\Omega}(x,t)=\{\overline{m},\alpha,\overline{u}\}$, since $\alpha$ is real. 

Thus, we will be dealing with a system of 10 PDEs (5 PDEs and their complex conjugates), linear and decoupled in the eigenfunctions but non-linear in the fields $\{\Omega,\overline\Omega\}$, with the exception of the $\{\chi,\psi\}$-coupling condition.

Let us consider a one-parameter Lie  group of infinitesimal transformations of the independent variables $\left\{x,t\right\}$, one dependent field and it complex conjugate $\left\{\Omega,\overline{\Omega}\right\}$, the spectral parameter $\lambda$, and the eigenfunctions $\left\{\chi,\overline{\chi},\psi,\overline{\psi}\right\}$, given by
\begin{equation}
    \begin{aligned}
    \tilde{x}&=x+\epsilon\,\xi_x(x,t,\lambda,\Omega,\overline{\Omega},\chi,\overline{\chi},\psi,\overline{\psi})\\
    \tilde{t}&=t+\epsilon\,\xi_t(x,t,\lambda,\Omega,\overline{\Omega},\chi,\overline{\chi},\psi,\overline{\psi})\\
    \tilde{\lambda}&=\lambda+\epsilon\,\xi_\lambda(x,t,\lambda,\Omega,\overline{\Omega},\chi,\overline{\chi},\psi,\overline{\psi})\\
    \tilde{\Omega}&=\Omega+\epsilon\,\eta_{\Omega}(x,t,\lambda,\Omega,\overline{\Omega},\chi,\overline{\chi},\psi,\overline{\psi})\\
    \tilde{\overline{\Omega}}&=\overline{\Omega}+\epsilon\,\eta_{\overline{\Omega}}(x,t,\lambda,\Omega,\overline{\Omega},\chi,\overline{\chi},\psi,\overline{\psi})\\
    \tilde{\chi}&=\chi+\epsilon\,\eta_\chi(x,t,\lambda,\Omega,\overline{\Omega},\chi,\overline{\chi},\psi,\overline{\psi})\\
    \tilde{\overline{\chi}}&=\overline{\chi}+\epsilon\,\eta_{\overline{\chi}}(x,t,\lambda,\Omega,\overline{\Omega},\chi,\overline{\chi},\psi,\overline{\psi})\\
    \tilde{\psi}&=\psi+\epsilon\,\eta_\psi(x,t,\lambda,\Omega,\overline{\Omega},\chi,\overline{\chi},\psi,\overline{\psi})\\
    \tilde{\overline{\psi}}&=\overline{\psi}+\epsilon\,\eta_{\overline{\psi}}(x,t,\lambda,\Omega,\overline{\Omega},\chi,\overline{\chi},\psi,\overline{\psi})\\
    \label{infinitesimal}
    \end{aligned}
\end{equation}
where $\epsilon$ is the group parameter. The associated vector field that generates the aforementioned infinitesimal transformation reads as
\begin{equation}\label{X}
X = \xi_x \frac{\partial}{\partial x} + 
\xi_t \frac{\partial}{\partial t}+
\xi_\lambda \frac{\partial}{\partial \lambda}+
\eta_\Omega \frac{\partial}{\partial \Omega}+
\eta_{\overline{\Omega}} \frac{\partial}{\partial \overline{\Omega}}+
\eta_{\chi} \frac{\partial}{\partial \chi}+
\eta_{\overline{\chi}} \frac{\partial}{\partial \overline{\chi}}+
\eta_{\psi} \frac{\partial}{\partial \psi}+
\eta_{\overline{\psi}} \frac{\partial}{\partial \overline{\psi}}
\end{equation}

This infinitesimal transformation induces a well known one in the derivatives of the fields and it  must preserve the invariance of the starting system of PDEs. By applying Lie's method \cite{BK,Olver,Lie}, this procedure yields an overdetermined system of PDEs for the infinitesimals called the determining equations, whose solutions provide the classical symmetries.  Lie symmetries have been independently computed with the help of Maple, Mathematica and Reduce. The main results are listed, for each case, in the following lines. 

\subsection{Lie symmetries for DNLS} \label{subsec: LSDNLS}
Classical Lie symmetries for the Lax pair \eqref{eq27}, its alternative Lax pair \eqref{2b} and its complex conjugates are
\begin{equation}
    \begin{aligned}
    \xi_x(x,t,\lambda,m,\overline{m},\chi,\overline{\chi},\psi,\overline{\psi})&=A_1x+A_2\\
    \xi_t(x,t,\lambda,m,\overline{m},\chi,\overline{\chi},\psi,\overline{\psi})&=2A_1t+A_3\\
    \xi_\lambda(x,t,\lambda,m,\overline{m},\chi,\overline{\chi},\psi,\overline{\psi})&=-A_1\,\lambda\\
    \eta_m(x,t,\lambda,m,\overline{m},\chi,\overline{\chi},\psi,\overline{\psi})&=\left(-\frac{A_1}{2}+iZ_1(t)\right)m\\
    \eta_{\overline{m}}(x,t,\lambda,m,\overline{m},\chi,\overline{\chi},\psi,\overline{\psi})&=\left(-\frac{A_1}{2}-iZ_1(t)\right)\overline{m}\\
    \eta_{\chi}(x,t,\lambda,m,\overline{m},\chi,\overline{\chi},\psi,\overline{\psi})&=K_1(\lambda)\,\chi\\
    \eta_{\overline{\chi}}(x,t,\lambda,m,\overline{m},\chi,\overline{\chi},\psi,\overline{\psi})&=\overline{K}_1(\lambda)\,\overline{\chi}\\
    \eta_{\psi}(x,t,\lambda,m,\overline{m},\chi,\overline{\chi},\psi,\overline{\psi})&=K_2(\lambda)\,\psi\\
    \eta_{\overline{\psi}}(x,t,\lambda,m,\overline{m},\chi,\overline{\chi},\psi,\overline{\psi})&=\overline{K}_2(\lambda)\,\overline{\psi}\\
    \end{aligned}
\end{equation}

where $A_i$, $i=1,...,3$ are arbitrary constants and $Z_1(t)$ is a real arbitrary function. $K_i(\lambda),\,\overline{K}_i(\lambda)$, $i=1,2$ are complex conjugate arbitrary functions of $\lambda$ that represent a phase shift in the eigenfunctions. 

\subsection{Lie symmetries for the modulus equation}
It may be noticed that in this case $\alpha$ represents a real field. Hence, Lie symmetries for Lax pairs \eqref{eq25} and \eqref{6b}, and their conjugates, read as
\begin{equation}
    \begin{aligned}
    \xi_x(x,t,\lambda,\alpha,\chi,\overline{\chi},\psi,\overline{\psi})&=A_1x+A_2\\
    \xi_t(x,t,\lambda,\alpha,\chi,\overline{\chi},\psi,\overline{\psi})&=2A_1t+A_3\\
    \xi_\lambda(x,t,\lambda,\alpha,\chi,\overline{\chi},\psi,\overline{\psi})&=-A_1\,\lambda\\
    \eta_\alpha(x,t,\lambda,\alpha,\chi,\overline{\chi},\psi,\overline{\psi})&=B_1\\
    \eta_{\chi}(x,t,\lambda,\alpha,\chi,\overline{\chi},\psi,\overline{\psi})&=K_1(\lambda)\,\chi\\
    \eta_{\overline{\chi}}(x,t,\lambda,\alpha,\chi,\overline{\chi},\psi,\overline{\psi})&=\overline{K}_1(\lambda)\,\overline{\chi}\\
    \eta_{\psi}(x,t,\lambda,\alpha,\chi,\overline{\chi},\psi,\overline{\psi})&=K_2(\lambda)\,\psi\\
    \eta_{\overline{\psi}}(x,t,\lambda,\alpha,\chi,\overline{\chi},\psi,\overline{\psi})&=\overline{K}_2(\lambda)\,\overline{\psi}\\
    \end{aligned}
\end{equation}

where $A_i, B_j$, $i=1,...,3; j=1$ are real arbitrary constants and $K_i(\lambda),\,\overline{K}_i(\lambda)$, $i=1,2$ are complex arbitrary functions of $\lambda$. 

\subsection{Lie symmetries for the decoupled $u$ equation}
Lie symmetries for the Lax pair \eqref{eq20}-\eqref{eq21} and their conjugates have been computed. In this case, we also need to take into account the coupling condition \eqref{eq9} for $u$ and $\overline{u}$ in the symmetry analysis. 
Thus, Lie symmetries for this case are given by
\begin{equation}
    \begin{aligned}
    \xi_x(x,t,\lambda,u,\overline{u},\chi,\overline{\chi},\psi,\overline{\psi})&=A_1x+2A_4t+A_2\\
    \xi_t(x,t,\lambda,u,\overline{u},\chi,\overline{\chi},\psi,\overline{\psi})&=2A_1t+A_3\\
    \xi_\lambda(x,t,\lambda,u,\overline{u},\chi,\overline{\chi},\psi,\overline{\psi})&=-A_1\,\lambda+A_4\\
    \eta_{u}(x,t,\lambda,u,\overline{u},\chi,\overline{\chi},\psi,\overline{\psi})&=\left(B_2+\frac{iA_4}{2}\right)x+Z_2(t)+iB_3\\
    \eta_{\overline{u}}(x,t,\lambda,u,\overline{u},\chi,\overline{\chi},\psi,\overline{\psi})&=\left(B_2-\frac{iA_4}{2}\right)x+Z_2(t)-iB_3\\
    \eta_{\chi}(x,t,\lambda,u,\overline{u},\chi,\overline{\chi},\psi,\overline{\psi})&=(-2iA_4t\lambda+K_1(\lambda))\,\chi\\
    \eta_{\overline{\chi}}(x,t,\lambda,u,\overline{u},\chi,\overline{\chi},\psi,\overline{\psi})&=(2iA_4t\lambda+\overline{K}_1(\lambda))\,\overline{\chi}\\
    \eta_{\psi}(x,t,\lambda,u,\overline{u},\chi,\overline{\chi},\psi,\overline{\psi})&=(2iA_4t\lambda+K_2(\lambda))\,\psi\\
    \eta_{\overline{\psi}}(x,t,\lambda,u,\overline{u},\chi,\overline{\chi},\psi,\overline{\psi})&=(-2iA_4t\lambda+\overline{K}_2(\lambda))\,\overline{\psi}\\
    \end{aligned}
\end{equation}
where we have used the same notation convention as before.

It is worth stressing the presence of a complete new symmetry associated to $A_4$, induced by the Miura transformation applied over $\alpha$, \eqref{eq6}.

In summary, classical Lie symmetries have been obtained for the three system considered above. These symmetries depend on a set of arbitrary parameters, listed as,
\begin{itemize}
\item The symmetries associated to the transformations of the independent variables $\left\{x,t,\lambda\right\}$ are expressed in terms of up to four real arbitrary constants $A_j$, $j=1,...,4$. Then, up to three additional real arbitrary constant $B_j$, $j=1,...,3$ may arise as symmetries for the fields, depending on each case.  
\item Two real arbitrary functions $Z_j(t), \,j=1,2$, which depend on the coordinate $t$.
\item Two complex arbitrary functions, $K_j(\lambda),\overline{K}_j(\lambda),\,j=1,2$,
appear in the transformations of the eigenfunctions of the Lax pairs.
\end{itemize}

Lie symmetries for Lax pairs generalize and include all the correspondent Lie symmetries for the starting non-linear PDEs. 

\section{Commutation relations and Lie Algebra}\label{sec:7}

In the ensuing sections, we will only consider the DNLS equation \eqref{eq0} and its Lax pairs \eqref{eq27} and \eqref{2b} for further analysis.  Hereafter, we will study the commutation relations among the infinitesimal generators and Lie algebra associated to the Lie symmetries for the Lax pairs for DNLS, displayed in subsection \ref{subsec: LSDNLS}. 

The eight resulting infinitesimal generators associated to these symmetries may be listed as
\begin{equation}
\begin{aligned}
&X_{1}={x}\frac{\partial}{\partial x}+2t\,\frac{\partial}{\partial t}-\lambda\frac{\partial}{\partial \lambda}-\frac{m}{2}\,\frac{\partial}{\partial m}-\frac{\overline{m}}{2}\,\frac{\partial}{\partial \overline{m}}\\
&X_{2}=\frac{\partial}{\partial x}\\
&X_{3}=\frac{\partial}{\partial t}\\
&Y_{\{Z_1(t)\}}=iZ_1(t)\left(m\,\frac{\partial}{\partial m}-\overline{m}\frac{\partial}{\partial \overline{m}}\right)\\
&\Gamma^{\chi}_{\{K_1(\lambda)\}}=K_1(\lambda)\,\chi\frac{\partial}{\partial \chi}\,,\qquad\qquad\qquad 
&\Gamma^{\overline{\chi}}_{\{\overline{K}_1(\lambda)\}}=\overline{K}_1(\lambda)\overline{\chi}\frac{\partial}{\partial \overline{\chi}}\\
&\Gamma^{\psi}_{\{K_2(\lambda)\}}=K_2(\lambda)\psi\frac{\partial}{\partial \psi}\,,\qquad\qquad\qquad
&\Gamma^{\overline{\psi}}_{\{\overline{K}_2(\lambda)\}}=\overline{K}_2(\lambda)\overline{\psi}\frac{\partial}{\partial \overline{\psi}}\\
\end{aligned}
\end{equation}
where generator $X_1-X_3$ arise exclusively from the arbitrary constants in the symmetries, $Y_{\{Z_1(t)\}}$ depends on an arbitrary function of time and $\Gamma^{\rho}_{\{\kappa(\lambda)\}}=\kappa(\lambda)\,\rho\partial_{\rho}$ denotes the generic generator associated to the arbitrary functions $\kappa(\lambda)=\{K_j(\lambda),\,\overline{K}_j(\lambda)\}$, $j=1,2$, and $\rho=\{\chi,\overline\chi,\psi,\overline\psi\}$.
\vspace{.2cm}

According to \cite{Steph}, symmetry generators depending on arbitrary constants will give rise to a Lie algebra, while generator depending on arbitrary functions will not, since we are dealing with an infinite-dimensional basis of generators. Notwithstanding this, the commutator of two symmetry generators is still a generator of a symmetry, written in terms of the involved arbitrary functions.

The commutation relations among these operators are presented in the following table
\begin{table}[H]
\centering
\resizebox{0.75\textwidth}{!} {
\begin{tabular}{c|cccccc}
      & $X_1$ & $X_2$ & $X_3$ & $Y_{\left\{Z_1(t)\right\}}$ & $\Gamma^{\rho}_{\left\{\kappa(\lambda)\right\}}$\\
	\hline
	$X_1$ & 0 & $-X_2$ & $-2X_3$ & $Y_{\left\{2t\,\frac{d Z_1}{dt}\right\}}$ & $\Gamma^{\rho}_{\left\{-\lambda \frac{d\kappa}{d\lambda}\right\}}$\\ 
	$X_2$ & $X_2$ & 0 & 0 & 0 & 0 \\ 
	$X_3$ & $2X_3$ & 0 & 0 & $Y_{\left\{\frac{d Z_1}{dt}\right\}}$ & 0\\ 
    $Y_{\left\{\tilde{Z}_1(t)\right\}}$ & $-Y_{\left\{2t\,\frac{d \tilde{Z}_1}{dt}\right\}}$ & 0 & $-Y_{\left\{\frac{d \tilde{Z}_1}{dt}\right\}}$ & 0 & 0\\ 
    $\Gamma^{{\rho}}_{\left\{\tilde{\kappa}(\lambda)\right\}}$ & $-\Gamma^{{\rho}}_{\left\{-\lambda \frac{d\tilde{\kappa}}{d\lambda}\right\}}$ & 0 & 0 & 0 & 0\\
    \end{tabular}}
\end{table}

Notice that the generic generator $\Gamma^{\rho}_{\{\kappa(\lambda)\}}$ defined above satisfy that\\ $\left[\Gamma^{{\rho}}_{\{\kappa(\lambda)\}},\Gamma^{\hat{\rho}}_{\{\hat{\kappa}(\lambda)\}}\right]=0$ for any combination of the arbitrary functions $\kappa(\lambda),\,\hat{\kappa}(\lambda)$ and the eigenfunctions $\rho,\,\hat{\rho}$. As mentioned, it may be observed that every commutator of two infinitesimal generators provides a non-trivial result, due to the presence of the arbitrary functions \cite{Olver}.

In general terms, these infinitesimal generators do not form a Lie algebra, but it is possible to obtain a finite-dimensional Lie algebra by adopting special values for the arbitrary functions, \cite{Champ87,David85}. 

In order to illustrate this, we study and classify the Lie algebra associated to simplest the ansatz $Z_1(t)=B_0$ constant, $K_1(\lambda)=K_2(\lambda)=0$. Thus, the following non-trivial commutation relations arise 
\begin{equation}
\left[X_1,X_2\right]=-X_2,\qquad \left[X_1,X_3\right]=-2X_3
\end{equation}
which may be identify as the four-dimensional real Lie algebra $\mathfrak{s}_{3,1}\oplus\mathfrak{n}_{1,1}$, with $a=\frac{1}{2}$, \cite{wint}.

\section{Similarity reductions}\label{sec:8}

Similarity reductions may be computed by solving the characteristic system 
\begin{equation} \label{characteristic}
\frac{dx}{\xi_{x}}=\frac{dt}{\xi_{t}}=\frac{d\lambda}{\xi_{\lambda}}=\frac{d\,m}{\eta_{m}}=\frac{d\,\overline{m}}{\eta_{\overline{m}}}=\frac{d\chi}{\eta_{\chi}}=\frac{d\overline{\chi}}{\eta_{\overline{\chi}}}=\frac{d\psi}{\eta_{\psi}}=\frac{d\overline{\psi}}{\eta_{\overline{\psi}}}
\end{equation}

The notation used for the reduced variables, reduced fields and reduced eigenfunctions is displayed as

\begin{table}[H]
\footnotesize
\centering
\resizebox{\textwidth}{!} {
\begin{tabular}{c c c}
\hline
& Original variables & New reduced variables\\
\hline\hline
Independent variables & $x, t, \lambda$ & $z,\,\Lambda$\\[.1cm]
Fields &$m(x,t), \overline{m}(x,t)$ & $\mathcal{M}(z), \overline{\mathcal{M}}(z)$\\[.1cm]

Eigenfunctions & $\chi(x,t,\lambda),\,\overline{\chi}(x,t,\lambda)$ & $ \Phi(z,\Lambda),\,\overline{\Phi}(z,\Lambda)$\\
& $\psi(x,t,\lambda),\,\overline{\psi}(x,t,\lambda)$ & $ \Psi(z,\Lambda),\,\overline{\Psi}(z,\Lambda)$\\
\hline
\end{tabular}
}
\label{Tab3}
\end{table}

The symmetries that will yield non-trivial reductions are those present in the transformations of the independent variables, \textit{i.e.}, the ones related to the arbitrary constants $A_i$, $i=1,...,3$. The rest of the symmetries will provide trivial reductions. Several reductions may emerge for different values of these constants, raising three independent reductions. 

Without loss of generality and for the sake of simplicity, we may consider all the arbitrary functions $K_j(\lambda),\,\overline{K}_j(\lambda)\,j=1,2$ to be zero, since the associated symmetries are  phase shifts over the eigenfunctions that are trivially satisfied due to the linearity of the Lax pairs. On the contrary, arbitrary functions depending on $t$, $Z_j(t),\,j=1,2$ do need to be taken into consideration.

Similarity reductions for the alternative spectral problem \eqref{2b} are displayed in \ref{app:RLP}.

\subsection{$A_1\neq 0$}\label{subsec:8.1} 

By solving the characteristic system \eqref{characteristic} in the general case, the following results have been obtained
\begin{itemize}
\item Reduced variable and reduced spectral parameter
\begin{equation}
z=\frac{A_1 x+ A_2}{\sqrt {A_1} \sqrt {A_3+2A_1t}}, \qquad 
\Lambda=\frac{\lambda}{\sqrt {A_1}}\sqrt {A_3+2A_1t}
\end{equation}

\item Reduced fields
\begin{equation}
\begin{aligned}
m(x,t)&=\mathcal{M}(z)\,\frac{A_1^\frac{1}{4}\,e^{i \int \frac{Z_1(t)}{A_3+2A_1t}dt}}{(A_3+2A_1t)^\frac{1}{4}}\\
\overline{m}(x,t)&=\overline{\mathcal{M}}(z)\,\frac{A_1^\frac{1}{4}\,e^{-i \int \frac{Z_1(t)}{A_3+2A_1t}dt}}{(A_3+2A_1t)^\frac{1}{4}}
\end{aligned}
\end{equation}

\item Reduced eigenfunctions
\begin{equation}
\begin{aligned}
\chi(x,t,\lambda)&=\Phi(z,\Lambda),\qquad&\overline{\chi}(x,t,\lambda)&=\overline{\Phi}(z,\Lambda)\\
\psi(x,t,\lambda)&=\Psi(z,\Lambda),\qquad &\overline{\psi}(x,t,\lambda)&=\overline{\Psi}(z,\Lambda)\\
\end{aligned}
\end{equation}

\item Reduced spectral problems
\begin{equation}
\begin{aligned}
&\Phi_{zz}-\left(i\Lambda +i\mathcal{M}\,\overline{\mathcal{M}} +\frac{\mathcal{M}_z}{\mathcal{M}}\right)\Phi_z-\frac{1}{4}\left(\mathcal{M}^2\,\overline{\mathcal{M}}^2+2i\mathcal{M} \overline{\mathcal{M}}_z\right)\Phi=0\\
&\Lambda\Phi_\Lambda-\left(\mathcal{M}\overline{\mathcal{M}}-\frac{i\mathcal{M}_z}{\mathcal{M}}+z-\Lambda\right)\Phi_z-\left(\frac{i}{4}\,\mathcal{M}^2{\overline{\mathcal{M}}}^2-\frac{1}{2}\mathcal{M}\overline{\mathcal{M}}_z-i\Lambda^2\right)\Phi=0
\end{aligned}
\label{PLPI}
\end{equation}
and its complex conjugate.

The alternative reduced $\Psi$-Lax pair can be found in \ref{app:RLP}, equation \ref{SLPI}.

\item Reduced Equation

The compatibility condition between both Lax pairs \eqref{PLPI} and \eqref{SLPI} and their complex conjugate provide the reduced equation (and its complex conjugate), which may be integrated as
\begin{equation}
\left[\frac{ i\mathcal{M}_{zz}}{\mathcal{M}}- \left(z +2  \mathcal{M} \overline{\mathcal{M}}\right)\frac{ \mathcal{M}_z}{\mathcal{M}}- \mathcal{M} \overline{\mathcal{M}}_z\right]_z =0
\label{REI}
\end{equation}

\end{itemize}

\subsection{$A_1=0$, $A_2\neq 0$, $A_3\neq 0$}\label{subsec:8.2}
By applying the same procedure, integrating \eqref{characteristic}, we get
\begin{itemize}
\item Reduced variable and reduced spectral parameter
\begin{equation}
z=\frac{A_2}{A_3}\left(x - \frac{A_2}{A_3} t\right), \qquad 
\Lambda=\frac{A_3}{A_2}\lambda
\end{equation}

\item Reduced fields
\begin{equation}
\begin{aligned}
m(x,t)&=\sqrt{\frac{A_2}{A_3}}\,e^{\frac{i}{A_3} \int Z_1(t)\,dt}\mathcal{M}(z)\\
\overline{m}(x,t)&=\sqrt{\frac{A_2}{A_3}}\,e^{-\frac{i}{A_3} \int Z_1(t)\,dt}\,\,\overline{\mathcal{M}}(z)
\end{aligned}
\end{equation}

\item Reduced eigenfunctions
\begin{equation}
\begin{aligned}
\chi(x,t,\lambda)&=\Phi(z,\Lambda),\qquad&\overline{\chi}(x,t,\lambda)&=\overline{\Phi}(z,\Lambda)\\
\psi(x,t,\lambda)&=\Psi(z,\Lambda),\qquad &\overline{\psi}(x,t,\lambda)&=\overline{\Psi}(z,\Lambda)\\
\end{aligned}
\end{equation}

\item Reduced spectral problems
\begin{equation}
\begin{aligned}
&\Phi_{zz}-\left(i\Lambda+i\mathcal{M} \overline{\mathcal{M}} +\frac{\mathcal{M}_z}{\mathcal{M}}\right)  \Phi_{z} -\frac{1}{4} \left(\mathcal{M}^2\overline{\mathcal{M}}^2+ 2i\mathcal{M} \overline{\mathcal{M}}_z\right) \Phi=0\\
&\left(1-\Lambda +\mathcal{M} \overline{\mathcal{M}}-i\frac{\mathcal{M}_z}{\mathcal{M}}\right)  \Phi_{z}+ \left(\frac{i}{4}\mathcal{M}^2\overline{\mathcal{M}}^2-\frac{1}{2}\mathcal{M} \overline{\mathcal{M}}_z-i \Lambda^2\right) \Phi=0
\end{aligned}
\label{PLPII}
\end{equation}
and its complex conjugate. 

The alternative reduced $\Psi$-Lax pair is displayed in \ref{app:RLP}, equation \ref{SLPII}.

\item Reduced Equations

The compatibility condition between both Lax pairs \eqref{PLPII} and \eqref{SLPII} and their complex conjugate provide the reduced equation 
\begin{equation}
\left[\frac{i\mathcal{M}_{zz}}{\mathcal{M}}- \left(1 +2  \mathcal{M} \overline{\mathcal{M}}\right)\frac{ \mathcal{M}_z}{\mathcal{M}}- \mathcal{M} \overline{\mathcal{M}}_z\right]_z =0
\end{equation}
\end{itemize}

\subsection{$A_1=0$, $A_2= 0$, $A_3\neq 0$}\label{subsec:8.3}
By integrating the characteristic system \eqref{characteristic}, the following results are obtained
\begin{itemize}
\item Reduced variable and reduced spectral parameter
\begin{equation}
z=x, \qquad 
\Lambda=\lambda
\end{equation}

\item Reduced fields
\begin{equation}
\begin{aligned}
m(x,t)&=e^{\frac{i}{A_3} \int Z_1(t)\,dt}\mathcal{M}(z)\\
\overline{m}(x,t)&=e^{-\frac{i}{A_3} \int Z_1(t)\,dt}\,\,\overline{\mathcal{M}}(z)
\end{aligned}
\end{equation}

\item Reduced eigenfunctions
\begin{equation}
\begin{aligned}
\chi(x,t,\lambda)&=\Phi(z,\Lambda),\qquad&\overline{\chi}(x,t,\lambda)&=\overline{\Phi}(z,\Lambda)\\
\psi(x,t,\lambda)&=\Psi(z,\Lambda),\qquad &\overline{\psi}(x,t,\lambda)&=\overline{\Psi}(z,\Lambda)\\
\end{aligned}
\end{equation}

\item Reduced spectral problems
\begin{equation}
\begin{aligned}
&\Phi_{zz}-\left(i\Lambda+i\mathcal{M} \overline{\mathcal{M}} +\frac{\mathcal{M}_z}{\mathcal{M}}\right)  \Phi_{z} -\frac{1}{4} \left(\mathcal{M}^2\overline{\mathcal{M}}^2+ 2i\mathcal{M} \overline{\mathcal{M}}_z\right) \Phi=0\\
&\left(-\Lambda +\mathcal{M} \overline{\mathcal{M}}-i\frac{\mathcal{M}_z}{\mathcal{M}}\right)  \Phi_{z}- \left(i \Lambda^2-\frac{1}{4}i\mathcal{M}^2\overline{\mathcal{M}}^2+\frac{1}{2}\mathcal{M} \overline{\mathcal{M}}_z\right) \Phi=0
\end{aligned}
\label{PLPIII}
\end{equation}
and its complex conjugate. 

The alternative reduced $\Psi$-Lax pair is written in \ref{app:RLP}, equation \ref{SLPIII}.

\item Reduced Equations

Finally, the reduce equation reads
\begin{equation}
\left[\frac{i\mathcal{M}_{zz}}{\mathcal{M}}- 2 \overline{\mathcal{M}}\mathcal{M}_z- \mathcal{M} \overline{\mathcal{M}}_z\right]_z =0
\end{equation}
\end{itemize}


\section{Conclusions}\label{sec:9}

In this paper we have analyzed some aspects of the integrability of the well-known derivative non-linear Schr\"{o}dinger equation in $1+1$ dimensions. This equation is presented as an integrable generalization of the famous non-linear Schr\"{o}dinger equation (NLS) with derivative-type non-linearity and it constitutes a differential equation of reference in the area of mathematical physics and soliton dynamics.

The Painlev\'e test has been proved to be a powerful technique to identify the integrability of this model, in combination with the application of a Miura transformation. The crucial aspect of our formulation lies in the Miura transformation, which allows us to connect the three differential equations and transform the original DNLS into a suitable PDE in $u$ where the Painlev\'e test with an unique branch of expansion is applicable. 

We have been able to successfully apply the singular manifold method in order to obtain two equivalent Lax pairs (up to a coupling constraint between the eigenfunctions involved) for three PDEs of interest derived from this procedure: the starting DNLS, a conservative PDE for the modulus $\alpha$ and a non-local Boussinesq-like equation for $u$.

Binary Darboux transformations are straightforward implemented, and easily yield the $\tau$-function and an iterative method to construct solutions. Rational soliton solutions have been obtained and their dynamics have been widely analyzed. 

We have determined the classical Lie symmetries for all these PDEs of interest and their Lax pairs. The main advantage of performing this procedure directly over the spectral problem is that it allows us to get simultaneously the symmetries related to the independent variables, fields and those associated to the eigenfunctions and the spectral parameter. Hence, Lie symmetries of the associated linear problem provide us more valuable information than the single analysis over the PDE itself. Three different set of symmetries have been obtained depending on up to seven arbitrary constant and five arbitrary functions of the independent variables $t$ or $\lambda$. 

The commutation relations among the associated generators have been studied and the Lie algebra has been identified for a particular choice of the arbitrary functions. Finally, we have analyzed three non-trivial similarity reductions in $1+1$ arising from the symmetries associated to the independent variables, where  the reduced equations and the reduced spectral problem have been simultaneously derived. 

\section*{Acknowledgements}
 This research
has been supported by MICINN (Grant PID2019-106820RB-C22) and
Junta de Castilla y Le\'on (Grant SA256P18). P. Albares also acknowledges support from the predoctoral grant FPU17/03246.

\newpage
\appendix

\section{Spectral problems in $1+1$}\label{app:LP}
In sections \ref{sec:2} and \ref{sec:3}, Lax pairs for the three main equations under study have been successfully derived. In this section we present the explicit results for the two equivalent Lax pairs ($\chi$-Lax pair and the alternative $\psi$-Lax pair) associated to each equation, which can be summarized as follows:

\subsection{Lax pairs for DNLS equation}
From \eqref{eq27} and properly combining equations \eqref{eq8}, \eqref{eq10} and \eqref{eq26} in \eqref{eq19a}, one gets the two equivalent spectral problems for the original DNLS equation \eqref{eq0},

\begin{subequations}
\begin{align}
    \label{2a}
    \chi_{xx}&=\chi_x\left[i\lambda+im{\overline{m}}+\frac{m_x}{m}\right]+\chi\left[\frac{1}{4}\,m^2{\overline{m}}^2+\frac{i}{2}m{\overline{m}_x}\right]\\ 
    \chi_{t}&=\chi_x\left[-\lambda+m{\overline{m}}-i\,\frac{m_x}{m}\right]+\chi\left[\frac{i}{4}\,m^2{\overline{m}}^2-\frac{1}{2}m{\overline{m}_x}-i\lambda^2\right]\nonumber\\[.5cm]
    \label{2b}
    \psi_{xx}&=\psi_x\left[-i\lambda-im{\overline{m}}+\frac{\left(m\,{\overline{m}}^2+2i{\overline{m}_x}\right)_x}{m\,{\overline{m}}^2+2i{\overline{m}_x}}\right]+\psi\left[\frac{1}{4}\,m^2{\overline{m}}^2+\frac{i}{2}\,m{\overline{m}_x}\right]\\
    \psi_{t}&=\psi_x\left[-\lambda+m{\overline{m}}+\frac{i\left(m\,{\overline{m}}^2+2i{\overline{m}_x}\right)_x}{m\,{\overline{m}}^2+2i{\overline{m}_x}}\right]+\psi\left[-\frac{i}{4}\,m^2{\overline{m}}^2+\frac{1}{2}\,m{\overline{m}_x}+i\lambda^2\right]\nonumber\\[.5cm]
    \label{2c}
    &\frac{\psi_x\chi_x}{\psi\chi}-\frac{1}{4}\,m{\overline{m}}\left(m{\overline{m}}+2i\,\frac{{\overline{m}_x}}{{\overline{m}}}\right)=0
\end{align}
     \label{LPDNLS1}
\end{subequations}

\subsection{Lax pairs for $\alpha(x,t)$} 
By an analogous procedure, the spectral problems associated to the equation for the modulus $\alpha(x,t)$ \eqref{eq4} read as
\begin{subequations}
\begin{align}
    \label{6a}
    \chi_{xx}&=\chi_x\left[i\lambda+\frac{i\alpha_x^2+\alpha_{xx}+i\alpha_t}{2\alpha_x}\right]+\chi\left[\frac{-\alpha_x^2+i\alpha_{xx}+\alpha_t}{2}\right]\\ 
    \chi_{t}&=\chi_x\left[-\lambda+\frac{\alpha_t-i\alpha_{xx}+\alpha_x^2}{2\alpha_x}\right]+\chi\left[-i\lambda^2+\frac{i\alpha_t-\alpha_{xx}-i\alpha_x^2}{2}\right]\nonumber\\[.5cm]
    \label{6b}
    \psi_{xx}&=\psi_x\left[-i\lambda+\frac{i\alpha_{x}^4+2i\alpha_{xxx}\alpha_x-2\alpha_x^2\alpha_{xx}-i\alpha_t^2-i\alpha_{xx}^2+2\alpha_{xt}\alpha_x}{2\alpha_x\left(-\alpha_x^2+i\alpha_{xx}+\alpha_t\right)}\right]\\
    &\,\,\,+\psi\left[\frac{-\alpha_x^2+i\alpha_{xx}+\alpha_t}{2}\right]\nonumber\\
    \psi_{t}&=\psi_x\left[-\lambda-\frac{\alpha_{x}^4+2\alpha_{xxx}\alpha_x+2i\alpha_x^2\alpha_{xx}-\alpha_t^2-\alpha_{xx}^2-2i\alpha_{xt}\alpha_x}{2\alpha_x\left(-\alpha_x^2+i\alpha_{xx}+\alpha_t\right)}\right]\nonumber\\
    &\,\,\,+\psi\left[i\lambda^2+\frac{-i\alpha_t+\alpha_{xx}+i\alpha_x^2}{2}\right]\nonumber\\[.5cm]
    \label{6c}
    &2\,\frac{\psi_x\chi_x}{\psi\chi}+\alpha_x^2-\,\alpha_t-i\,\alpha_{xx}=0
\end{align}
\label{alphaLP}
\end{subequations}

\subsection{Lax pairs for $u(x,t)$}
Regarding equations \eqref{eq19}-\eqref{eq21}, the spectral problems for $u$ may be written as
\begin{subequations}
\begin{align}
    \label{11a}
    \chi_{xx}&=\chi_x\left[i\lambda+\frac{u_{xxx}+iu_{xt}}{2u_{xx}}\right]-u_{xx}\chi\\ 
    \chi_{t}&=\chi_x\left[-\lambda+\frac{-iu_{xxx}+u_{xt}}{2u_{xx}}\right]-i\left[\lambda^2+u_{xx}\right]\chi\nonumber\\[.5cm]
    \label{11b}
    \psi_{xx}&=\psi_x\left[-i\lambda+\frac{u_{xxx}-iu_{xt}}{2u_{xx}}\right]-u_{xx}\psi\\
    \psi_{t}&=\psi_x\left[-\lambda+\frac{iu_{xxx}+u_{xt}}{2u_{xx}}\right]+i\left[\lambda^2+u_{xx}\right]\psi\nonumber\\[.5cm]
    \label{11c}
    &\frac{\psi_x\chi_x}{\psi\chi}+u_{xx}=0
\end{align}
\label{DUWLP}
\end{subequations}

\section{Reduced spectral problems}\label{app:RLP}
In the present section we display the explicit computation of the reduced spectral problems $\Psi$-Lax pairs. These Lax pairs arise from the similarity reductions performed over the alternative $\psi$-Lax pairs in each case of study in section \ref{sec:8}. The compatibility condition over the reduced Lax pairs in all cases yields
\begin{equation}
\frac{\Phi_{z}\Psi_{z}}{\Phi \Psi}-\frac{1}{4}\mathcal{M}\left(\mathcal{M}{\overline{\mathcal{M}}}^2+2i\overline{\mathcal{M}}_z\right)=0\nonumber
\end{equation}

\subsection{$\Psi$-Lax pair for Case \ref{subsec:8.1}}
\begin{equation}
\begin{aligned}
\Psi_{zz}&-\frac{1}{4}\mathcal{M}\left(\mathcal{M}{\overline{\mathcal{M}}}^2+2i{\overline{\mathcal{M}}_z}\right)\Psi\\
&+i\left(\Lambda+\mathcal{M}\overline{\mathcal{M}}+\frac{\left(i\mathcal{M}\overline{\mathcal{M}}^2-2\overline{\mathcal{M}}_z\right)_z}{\mathcal{M}\overline{\mathcal{M}}^2+2i\overline{\mathcal{M}}_z}\right)\Psi_z=0\\[.2cm]
\Lambda\Psi_\Lambda&+\left(\Lambda-z-\mathcal{M}\overline{\mathcal{M}}-\frac{\left(i\mathcal{M}\overline{\mathcal{M}}^2-2\overline{\mathcal{M}}_z\right)_z}{\mathcal{M}\overline{\mathcal{M}}^2+2i\overline{\mathcal{M}}_z}\right)\Psi_z\\
&+\frac{i}{4}\left(\mathcal{M}^2{\overline{\mathcal{M}}}^2+2i\mathcal{M}\overline{\mathcal{M}}_z-4\Lambda^2\right)\Psi=0
\end{aligned}
\label{SLPI}
\end{equation}
and its complex conjugate.

\subsection{$\Psi$-Lax pair for Case \ref{subsec:8.2}}
\begin{equation}
\begin{aligned}
\Psi_{zz}&-\frac{1}{4}\mathcal{M}\left(\mathcal{M}{\overline{\mathcal{M}}}^2+2i{\overline{\mathcal{M}}_z}\right)\Psi\\
&+i\left(\Lambda+\mathcal{M}\overline{\mathcal{M}}+\frac{\left(i\mathcal{M}\overline{\mathcal{M}}^2-2\overline{\mathcal{M}}_z\right)_z}{\mathcal{M}\overline{\mathcal{M}}^2+2i\overline{\mathcal{M}}_z}\right)\Psi_z=0\\[.2cm]
&\left(\Lambda-1-\mathcal{M}\overline{\mathcal{M}}-\frac{\left(i\mathcal{M}\overline{\mathcal{M}}^2-2\overline{\mathcal{M}}_z\right)_z}{\mathcal{M}\overline{\mathcal{M}}^2+2i\overline{\mathcal{M}}_z}\right)\Psi_z\\
&+\frac{i}{4}\left(\mathcal{M}^2{\overline{\mathcal{M}}}^2+2i\mathcal{M}\overline{\mathcal{M}}_z-4\Lambda^2\right)\Psi=0
\end{aligned}
\label{SLPII}
\end{equation}
and its complex conjugate.

\subsection{$\Psi$-Lax pair for Case \ref{subsec:8.3}}
\begin{equation}
\begin{aligned}
\Psi_{zz}&-\frac{1}{4}\mathcal{M}\left(\mathcal{M}{\overline{\mathcal{M}}}^2+2i{\overline{\mathcal{M}}_z}\right)\Psi\\
&+i\left(\Lambda+\mathcal{M}\overline{\mathcal{M}}+\frac{\left(i\mathcal{M}\overline{\mathcal{M}}^2-2\overline{\mathcal{M}}_z\right)_z}{\mathcal{M}\overline{\mathcal{M}}^2+2i\overline{\mathcal{M}}_z}\right)\Psi_z=0\\[.2cm]
&\left(\Lambda-\mathcal{M}\overline{\mathcal{M}}-\frac{\left(i\mathcal{M}\overline{\mathcal{M}}^2-2\overline{\mathcal{M}}_z\right)_z}{\mathcal{M}\overline{\mathcal{M}}^2+2i\overline{\mathcal{M}}_z}\right)\Psi_z\\
&+\frac{i}{4}\left(\mathcal{M}^2{\overline{\mathcal{M}}}^2+2i\mathcal{M}\overline{\mathcal{M}}_z-4\Lambda^2\right)\Psi=0
\end{aligned}
\label{SLPIII}
\end{equation}
and its complex conjugate.

\end{document}